\documentclass[a4paper]{article}
\usepackage{graphics,setspace,amsmath}
\usepackage{natbib}

\newcommand{\ket}[2]{|#1\rangle _{#2}}
\newcommand{\bra}[2]{\langle _{#1}#2|}
\newcommand{\up}{\!\uparrow}
\newcommand{\down}{\!\downarrow}

\begin{document}

\begin{center}\LARGE Philosophical Aspects of Quantum Information Theory\footnote{August 2006. Chapter for D. Rickles (ed.) \textit{The Ashgate Companion to the New Philosophy of Physics} (Ashgate, forthcoming 2007). Email: \texttt{c.g.timpson@leeds.ac.uk}}
\end{center}
\begin{center}
\large Christopher G. Timpson\\{\small \textit{Division of History and Philosophy of Science,}}\\{\small \textit{Department of Philosophy, University of Leeds, LS2 9JT, UK.}}\end{center}

\tableofcontents

\newpage
\section{Introduction}

While quantum information theory is one of the most lively, up-and-coming new areas of research in physics, its central concerns have long been familiar. They are simply those that have lain close to the heart of anyone interested in the foundations of quantum mechanics, since its inception: How does the quantum world differ from the classical one? 

What is distinctive about the field, however, is that this question is approached from a particular viewpoint: a task-oriented one. It has turned out to be most productive to ask: what can one \textit{do} with quantum systems, that one could not with classical ones? What use can one make of non-commutativity, entanglement; and the rest of our familiar friends? 

\sloppy The answers have involved identifying a rich range of communication and computational tasks that are distinctively quantum mechanical in nature: notions, for example, of quantum computation, quantum cryptography and entanglement-assisted communication. Providing these answers has  deepened our understanding of quantum theory considerably, while spurring impressive experimental efforts to manipulate and control individual quantum systems. What is surprising, and, \textit{prima facie}, need not have been the case, is that the peculiar behaviour of quantum systems does provide such interesting opportunities for new forms of communication and computation, when one might have feared that these peculiarities would only present annoying obstacles for the increasing miniaturisation of information processing devices.   

For philosophers, and for those interested in the foundations of quantum mechanics, quantum information theory therefore makes a natural and illuminating object of study. There is a great deal to be learnt therein about the behaviour of quantum systems that one did not know before. We shall survey a few of these points here. But there are further reasons why quantum information theory is particularly intriguing. 

Running along with the development of the field have been a number of more-or-less explicitly philosophical propositions. Many have felt, for example, that the development of quantum information theory heralds the dawn of a new phase of physical theorising, in which the concept of information will come to play a much more fundamental r\^{o}le than it has traditionally been assigned. Some have gone so far as to re-vivify immaterialist ideals by arguing that information should be seen as the basic category from which all else flows, and that the new task of physics will be to describe how this information evolves and manifests itself. \citet{wheeler} is the cheerleader for this sort of view. Or again, the rallying cry of the quantum information scientist is that `Information is Physical!', a doctrine of surprising-sounding ontological import. On the less extreme side is the widespread view that developments in quantum information will finally help us sort out the conceptual problems in quantum mechanics that have so vexed the theory from the beginning.  

In order to get clearer on what import quantum information theory \textit{does} have, it would be beneficial to gain a better understanding of what the theory is about. This will be one of our main aims here. In Section~\ref{first} we will survey some elementary aspects of quantum information theory, with a focus on some of the principles and heuristics involved. In Section~\ref{info} we will examine in detail what exactly quantum information (and therefore quantum information \textit{theory}) is; and deploy our findings in resolving puzzles surrounding the notion of quantum teleportation. This will provide us with a better grasp of the relation between information theory and the world. In Section~\ref{physicalside} 
we turn to examine what one might learn from the development of quantum computation, both about quantum systems and about the theory of computation, asking where the speed-up in quantum computers might come from and what one should make of the Church-Turing hypothesis in this new setting. Finally, in Section~\ref{foundations}, we broach the compelling question of what, if anything, quantum information theory might have to teach us about the traditional foundational problems in quantum mechanics. Some pit-falls are noted before we discuss a number of attempts to provide information-theoretic axiomatisations of quantum mechanics: Zeilinger's Foundational Principle, the CBH theorem and quantum Bayesianism. On all of these matters there is more to be said than I essay here.

In general there are two kinds of strategies that have been manifest in attempts to obtain philosophical or foundational dividends from quantum information theory, the direct and the indirect. We will canvass a number of each. The direct strategies include such thoughts as these: the quantum state is to be understood as information; quantum information theory supports some form of immaterialism; quantum computation is evidence for the Everett interpretation. None of these survives close examination, and it seems unlikely that any such direct attempt to read a philosophical lesson from quantum information theory will. Much more interesting and substantial are the indirect approaches which seek, for example, to learn something useful about the structure or axiomatics of quantum theory by reflecting on quantum information-theoretic phenomena; that might look to quantum information theory to provide new analytic tools for investigating that structure; or that look to suggested constraints on the power of computers as potential constraints on new physical laws. The deepest lessons are perhaps still waiting to be learnt.

\section{First steps with quantum information}\label{first}

As I have said, 
quantum information theory is animated by the thought that the difference in character of quantum and classical systems makes possible interesting new forms of communication and computation. And one may reasonably hope that reflecting on the nature and possibility of these new tasks will in turn shed light back on the differences between quantum and classical.
Quantum information theory may be seen as an extension of classical information theory that introduces new primitive information-theoretic resources, particularly \textit{quantum} bits and shared entanglement; and develops quantum generalisations of the associated notions of sources, channels and codes. Within this general setting, one may then devise cryptographic, communication or computational tasks that go beyond the classical, and investigate their properties.

\subsection{Bits and qubits}

It is useful to begin by focusing on the differences between the familiar classical primitive---the bit---and the corresponding quantum primitive---the \textit{qubit} (quantum bit)\footnote{The term `qubit' was introduced in print in \citet{qcoding}, the concept having been first aired by Schumacher, following conversations with Wootters, at the IEEE meeting on the Physics of Computation in Dallas, October 1992.}. A classical bit is some physical object which can occupy one of two distinct, stable classical states, conventionally labelled by the binary values 0 or 1. The term `bit' is also used to signify an \textit{amount} of classical information: the number of bits that would be required to encode the output of a source is called the quantity of information the source produces \citep{shannon}. We shall see more of this below (Section~{\ref{info}}).      

A qubit is the precise quantum analogue of a bit: it is a two-state \textit{quantum} system. Examples might be the spin degree of freedom of an electron or of a nucleus, or an atom with an excited and an unexcited energy state, or the polarization of a photon. The two basic orthogonal states of a qubit are represented by vectors labelled $\ket{0}{}$ and $\ket{1}{}$. These states are called the \textit{computational basis} states and provide analogues of the classical 0 and 1 states. But of course, analogy is not identity. While a classical bit may only exist in either the 0 or 1 states, the same is not true of a qubit. \textit{It} may exist in an arbitrary superposition of the computational basis states: $\ket{\psi}{} = \alpha \ket{0}{} + \beta \ket{1}{}$, where $\alpha$ and $\beta$ are complex numbers whose moduli squared sum to one. There are, therefore, continuously many different states that a qubit may occupy, one for each of the different values the pair $\alpha$ and $\beta$ may take on; and this leads to the natural thought that qubits contain vastly more information than classical bits, with their measly two element state space. Intuitively, this enormous difference in the amounts of information associated with bit and qubit might seem to be their primary information-theoretic distinction. 

However a little care is required here. While it is certainly true that the existence of superpositions represents a fundamental difference between qubits and bits, it is not straightforward to maintain that qubits therefore contain vastly more information. For a start, it is only under certain conditions that systems may usefully be said to contain information at all---typically only when they are playing a suitable role in a communication protocol of some sort. But more importantly, we need to make a distinction between two different notions of information that coincide in the classical case, but diverge in the quantum; that is, a distinction between \textit{specification} information and \textit{accessible} information.  

Consider a sequence of $N$ systems, each of which has been prepared in some particular state from a given finite set of states (the very simplest case would be a sequence of bits which has been prepared in some sequence of 0s and 1s). Assume, furthermore, that each particular state occurs in the sequence with a given probability. We may think of this sequence as being our message. 

We may now ask how much information (in bits) is required to specify what this sequence of states is. This is called the specification information associated with the message. We might also ask how much information can be acquired or read from the sequence: this is the accessible information. Clearly, in the classical case, the two quantities will coincide, as classical states are perfectly distinguishable. When presented with the message, we may determine the sequence of states perfectly by observation or measurement; and what we have determined---the identity of the sequence of states the message comprises---evidently gives us enough information to specify what that sequence is. However, in the quantum case, these two quantities will differ, in general. If we prepare our $N$ systems in a sequence of states drawn from a set of non-orthogonal quantum states, it will \textit{not} be possible to identify the sequence of states by measurement. This means that in general much more information will be required to specify the sequence than may be obtained from it. Take the case of a sequence of qubits. As we have said, there are continuously many states that each qubit could be prepared in, so the specification information associated with the sequence could be unboundedly large. But it would only be if each of the qubits were prepared in one or other of two fixed \textit{orthogonal} states that we could reliably identify what the sequence of states prepared actually was; and then we would only be getting one bit of information per qubit in the sequence. 

It turns out that this would in fact be the best that we could do. A striking result due to \citet{holevo}, called the Holevo bound, establishes that the maximum amount of information that can be obtained from measurements on a quantum system is given by the logarithm (to base 2) of the number of orthogonal states the system possesses, no matter how clever our measuring procedure. Thus, in the case of qubits, the maximum amount of information per qubit that can be decoded from measurements on the sequence is just one bit. Given that `encoded' is a success word (one can't be said to have encoded something if one cannot, in principle \textit{decode} it), this tells us that the maximum amount of information that can be encoded into a qubit is just one bit; the same amount, of course, as a classical bit. So while we may prepare some sequence of qubits having an unboundedly large specification information, we could not thereby have managed to encode more than a \textit{single} bit of information into each qubit. Looked at from a certain perspective, this presents an intriguing puzzle. As Caves and Fuchs have put it: just why is the state-space of quantum mechanics so gratuitously large, from the point of view of storing information? \citep{fuchs:howmuch}.       

There is a final important reason why we should not, on reflection, have been tempted to conclude that qubits can contain vastly more information than classical bits, on the strength of the possibility of preparing them in superpositions of computational basis states. It is that the intuition driving this thought derives from an overly classical way of thinking about and quantifying information. If we could prepare a classical system in any one of an arbitrarily large number of different states, then it might indeed be appropriate to associate an arbitrarily large amount of information with that system. Classical information. But quantum systems are not classical systems and quantum states are not classical states. It was Schumacher's insight \citep{qcoding} that this allowed us to introduce a new notion of information peculiar to quantum systems---quantum information. And we need a \textit{new} theory to tell us how much of \textit{this} information there may be about in a given situation (we will see how Schumacher developed this in Section~\ref{info}). Thus when talking about the amount of information that is associated with a given system, or has been encoded into it, we need to clarify whether we are talking about transmitting \textit{classical} information \textit{using} quantum systems, or whether we are talking about encoding and transmitting quantum information properly so-called. In the former context, the notions of specification and accessible information apply: how much classical information is required to specify a sequence, or how much classical information one can gain from it, respectively; and we know that at most one classical bit can be encoded into a qubit. In the latter context, we apply the appropriate measure of the amount of quantum information; and it may come as no surprise to learn that the maximum amount of quantum information that may be encoded into a qubit is one qubit's worth! (See below.)  

\subsection{The no-cloning theorem}

The difference in the nature of the state spaces of bit and qubit---the fact that qubits can support superpositions and hence enjoy a large number of distinct, but non-distinguishable states---does not, therefore, manifest itself in a simple-minded difference in the amount of information the two types of objects can contain, but in more subtle and interesting ways. We have already seen one, in the ensuing difference between accessible and specification information. A closely related idea is that of \textit{no-cloning}.  

We have already used the idea that it is not possible to distinguish perfectly between non-orthogonal quantum states; equivalently, that it is not possible to determine an unknown state of a single quantum system. If we don't at least know an orthogonal set the state in question belongs to (e.g., the basis the system was prepared in) then no measurement will allow us to find out its state reliably\footnote{Imagine trying to determine the state by measuring in some basis. One will get some outcome corresponding to one of the basis vectors. But was the system actually \textit{in} that state before the measurement? Only if the orthogonal basis we chose to measure in was one containing the unknown state. And even if we happened on the right basis by accident, we couldn't \textit{know} that from the result of the measurement, so we could not infer the identity of the unknown state. For a fully general discussion, see \citet{busch:observable}.}. This result is logically equivalent to an important constraint on information processing using quantum systems. 

Whether we are primarily concerned with encoding classical information or quantum information into quantum systems, we will be involved in preparing those systems in various quantum states. The no-cloning theorem due to \citet{dieks} and \citet{wootters:zurek} states that it is impossible to make \textit{copies} of an unknown quantum state. Presented with a system in an unknown state $\ket{\psi}{}$, there is no way of ending up with more than one system in the same state $\ket{\psi}{}$. One can swap $\ket{\psi}{}$ from one system to another\footnote{Take two Hilbert spaces of the same dimension, ${\cal{H}}_{1}$ and ${\cal{H}}_{2}$. The `swap' operation $U_{\textrm{S}}$ on ${\cal{H}}_{1}\otimes{\cal{H}}_{2}$ is a unitary operation that swaps the state of system 1 for the state of system 2 and \textit{vice versa}: $U_{\textrm{S}}\ket{\psi}{1}\ket{\psi^{\prime}}{2} = \ket{\psi^{\prime}}{1}\ket{\psi}{2}$. If we take $\{\ket{\phi_{i}}{1,2}\}$ as basis sets for ${\cal{H}}_{1}$ and ${\cal{H}}_{2}$ respectively, then $U_{\textrm{S}} = \sum_{ij}\ket{\phi_{j}}{1\,1}\!\bra{}{\phi_{i}}\otimes   \ket{\phi_{i}}{2\,2}\!\bra{}{\phi_{j}}$, for example.}, but one can't copy it.
This marks a considerable difference from classical information processing protocols, as in the classical case, the value of a bit may be freely copied into numerous other systems, perhaps by measuring the original bit to see its value, and then preparing many other bits with this value. The same is not possible with quantum systems, obviously, given that we can't determine the state of a single quantum system by measurement: the measuring approach would clearly be a non-starter.

To see that no more general scheme would be possible either, consider a device that makes a copy of an unknown state $\ket{\alpha}{}$. This would be implemented by a unitary evolution\footnote{Is it too restrictive to consider only unitary evolutions? One can always consider a non-unitary evolution, e.g. measurement, as a unitary evolution on a larger space. Introducing auxiliary systems, perhaps including the state of the apparatus, doesn't affect the argument.}  $U$ that takes the product $\ket{\alpha}{}\ket{\psi_{0}}{}$, where $\ket{\psi_{0}}{}$ is a standard state, to the product $\ket{\alpha}{}\ket{\alpha}{}$. Now consider another possible state $\ket{\beta}{}$. Suppose the device can copy this state too: $U\ket{\beta}{}\ket{\psi_{0}}{}=\ket{\beta}{}\ket{\beta}{}$. If it is to clone a general unknown state, however, it must be able to copy a superposition such as $\ket{\xi}{}=1/\sqrt{2}(\ket{\alpha}{}+\ket{\beta}{})$ also, but the effect of $U$ on $\ket{\xi}{}$ is to produce an entangled state  $1/\sqrt{2}(\ket{\alpha}{}\ket{\alpha}{} + \ket{\beta}{}\ket{\beta}{})$ rather than the required $\ket{\xi}{}\ket{\xi}{}$. It follows that no general cloning device is possible. This argument makes use of a central feature of quantum dynamics: its \textit{linearity}\footnote{An operator $O$ on $\cal{H}$ is linear if its effect on a linear combination of vectors is equal to the same linear combination of the effects of the operator on each vector taken individually: $O(\alpha \ket{u_{1}}{} + \beta \ket{u_{2}}{}) = \alpha \,O\ket{u_{1}}{} + \beta O\ket{u_{2}}{} = \alpha\ket{v_{1}}{} + \beta\ket{v_{2}}{}; \ket{u_{i}}{}, \ket{v_{i}}{} \in {\cal{H}}$. Unitary operators are, of course, linear.}.

In fact it may be seen in the following way that if a device can clone more than one state, then these states must belong to an orthogonal set. We are supposing that $U\ket{\alpha}{}\ket{\psi_{0}}{}=\ket{\alpha}{}\ket{\alpha}{}$ and $U\ket{\beta}{}\ket{\psi_{0}}{}=\ket{\beta}{}\ket{\beta}{}$. Taking the inner product of the first equation with the second implies that 
$ \langle\alpha|\beta\rangle=\langle\alpha|\beta\rangle^{2},$
which is only satisfied if $\langle\alpha|\beta\rangle=0 \text{ or } 1$, i.e., only if $\ket{\alpha}{}$ and $\ket{\beta}{}$ are identical or orthogonal.

I said above that no-cloning was logically equivalent to the impossibility of determining an unknown state of a single system. We have already seen this in one direction: if one could determine an unknown state, then one could simply do so for the system in question and then construct a suitable preparation device to make as many copies as one wished, as in the classical measuring strategy. What about the converse? If one could clone, could one determine an unknown state? The answer is yes. If we are given sufficiently many systems all prepared in the same state, then the results of a suitable variety of measurements on this group of systems will furnish one with knowledge of the identity of the state (such a process is sometimes called \textit{quantum state tomography}). For example, if we have a large number of qubits all in the state $\ket{\psi}{} = \alpha\ket{0}{} + \beta\ket{1}{}$, then measuring them one by one in the computational basis will allow us to estimate the Born rule probabilities $|\langle 0|\psi\rangle|^{2} = |\alpha|^{2}$ and $|\langle 1|\psi\rangle|^{2} = |\beta|^{2}$, with increasing accuracy as the number of systems is increased. This only gives us some information about the identity of $\ket{\psi}{}$, of course. To determine this state fully, we also need to know the relative phase of $\alpha$ and $\beta$. One could find this by also making a sufficient number of measurements on further identically prepared individual systems in the rotated bases $\{1/\sqrt{2}(\ket{0}{}\pm\ket{1}{}\}$ and $\{1/\sqrt{2}(\ket{0}{}\pm i\ket{1}{}\}$, for example \citep{fano,park:band}. (One would need to make more types of measurement if the system were higher dimensional. For an $n$-dimensional system, one needs to establish the expectation values of a minimum of $n^{2}-1$ operators.) Thus access to many copies of identically prepared systems allows one to find out their state; and with a cloner, one could multiply up an individual system into a whole ensemble all in the same state; so cloning would allow identification of unknown states. (It would also imply, therefore, the collapse of the distinction between accessible and specification information.)       

In fact it was in the context of state determination that the question of cloning first arose \citep{herbert:1982}. Cloning would allow state determination, but then \textit{this} would give rise to the possibility of superluminal signalling using entanglement in an EPR-type setting: one would be able to distinguish between different preparations of the same density matrix, hence determine superluminally which measurement was performed on a distant half of an EPR pair. The no-cloning theorem was derived to show that this possibility is ruled out.

So the no-cloning theorem is not only interesting from the point of view of showing differences between classical and quantum information processing, important as that is. It also illustrates in an intriguing way how tightly linked together various different aspects of the quantum formalism are. The standard proof of no-cloning is based on the fundamental linearity property of the dynamics: suggestive if one were searching for information-theoretic principles that might help illuminate aspects of the quantum formalism. Furthermore, cloning is logically equivalent to the possibility of individual state determination and hence implies superluminal signalling; thus no-cloning seems to be a crucial part of the apparent peaceful co-existence between quantum mechanics and relativity. All this might seem to suggest some link between no-signalling and linearity of the dynamics: see \citet{svetlichny:1998} and \citet{simon:no-signalling} for some work in this connection (but cf. \citet{svetlichny:comment} also); \citet{horodecki:common-origin} discuss no-cloning and the related idea of no-deleting in a general setting.

\subsection{Quantum cryptography} 

Quantum cryptography is the study of the possibilities of secret communication using quantum properties. It holds out the promise of security of communication guaranteed by the laws of physics, in contrast to the mere computational difficulty that underwrites our best in classical security. In doing so it makes essential use of the fact that non-orthogonal quantum states cannot be perfectly distinguished; essential use, that is, of the great size of the qubit state space that, in a sense, we have seen we lack access to. The existence of non-orthogonal states is linked, of course, to the non-commutativity of observables and the existence of incompatible physical quantities. One of the reasons, therefore, that quantum cryptography has been of interest is that it provides a very direct `cash-value' practical application of---and new theoretical playground for---some of the most puzzling and non-classical aspects of the quantum formalism\footnote{For example, the study of quantum cryptography has provided very useful conceptual and formal tools for clarifying and quantifying what had been the unsatisfactorily messy matters of what, if anything, measurement and disturbance have to do with one another. The folklore, since Heisenberg, has not been edifying. See \citet{fuchs:info-gain} and \citet{fuchsjacobs:info-trade}. The lesson is to focus on \textit{states}; and non-orthogonality is the crucial thing. Measurements disturb non-orthogonal sets of states, but if a state is known to be from some \textit{orthogonal} set, it is, perhaps surprisingly, possible to measure \textit{any} observable on it you wish and return it to its initial state, i.e., to leave it undisturbed.}. 

How might one go about using qubits for secret communication? One thought might be to try to hide the secret message directly in a sequence of qubits (this was the form that one of the very earliest protocols in fact took \citep{BBB:82,brassard:history}). So, for example, one party, Alice, might encode a classical message (a sequence of 0s and 1s, say) into a sequence of quantum systems by preparing them in various non-orthogonal states. Thus spin-up and spin-down might represent 0 and 1 respectively; and for each qubit in her sequence, she could choose what basis to prepare it in. Picking from $\sigma_{z}$ and $\sigma_{x}$ bases, for example, her encoded message will be an alternating sequence of $\sigma_{z}$ and $\sigma_{x}$ eigenstates, with the eigenvalue of each indicating the classical bit value encoded. So a sequence like

\[ \ket{\up_{z}}{}\ket{\down_{z}}{}\ket{\down_{x}}{}\ket{\up_{x}}{}\ket{\up_{z}}{}\; \mathrm{would\;\; represent\; the \; message}\;\; 01100.\] 

Now if the other party, Bob, for whom the message is intended, knows what sequence of bases Alice chose---that is, if they have met previously and agreed upon the basis sequence clandestinely---then he is able to measure in the appropriate basis for each system and read out correctly what the classical bit value encoded is. However, any eavesdropper, Eve, who wishes to learn the message, cannot do so, as she doesn't know which basis each system was prepared in. All she can have access to is a sequence of non-orthogonal states; and we know that she will be unable to identify what that sequence of states is; therefore she will be unable to learn the secret message. Furthermore, if she does try to learn something about the identity of the sequence of states, she will end up disturbing them in such a way that Alice and Bob will be able to detect her eavesdropping. They will then know that if they wish to preserve the security of future transmissions they will need to meet once more and agree upon a new secret sequence of encoding bases. If there is no eavesdopping, though, they may keep on using the same encoding basis sequence over and over again.

However it turns out that this sort of protocol isn't the best one to use. Although Eve cannot fully identify the sequence of non-orthogonal states---and hence the secret message---by measurement, she will be able to gain some information about it\footnote{It is for this reason that Alice and Bob would have to change their agreed basis sequence after detecting the presence of Eve. If they didn't then Eve would eventually be able to gain enough information about the encoding basis sequence to learn a good deal about the messages being sent.}; and her actions in trying to gather information will end up scrambling some of the message that Alice is trying to send Bob---he will not receive everything that Alice is trying to send. One can avoid these kinds of problems and generate a perfectly secure protocol by making use of the ideas of \textit{key distribution} instead \citep{BB:84}.

\subsubsection{Key Distribution}
There are two central techniques here, both developed before the advent of quantum cryptography. The first is called \textit{symmetrical} or \textit{private-key} cryptography; the second, \textit{asymmetrical} or \textit{public-key} cryptography. In both techniques the message being sent is encrypted and rendered unreadable using a key---and a key is required to unlock the message and allow reading once more.

In private-key cryptography, both parties share the same, secret, key, which is used both for encryption and decryption. The best known (and the only known provably secure) technique is the \textit{one-time pad}. Here the key consists of a random string of bit values, of the same length as the message to be encrypted. The message string is encrypted simply by adding (modulo 2) the value of each bit in the message to the value of the corresponding bit in the key string. This generates a cryptogram which is just as random as the bit values in the private key and will thus provide Eve with no information about the message. The cryptogram is decrypted by subtracting (again modulo 2) the key from the cryptogram, returning the starting message string. Thus if Alice and Bob share a random secret key, they can communicate securely. The down-side to this protocol is that each key may only be used once. If more than one message were encoded using the same key then Eve could begin to identify the key by comparing the cryptograms. Also, whenever Alice and Bob wish to share a new key, they must meet in secret, or use a trusted courier; and a key has to be as long as any message sent. Hence the preference for public-key cryptography in the majority of cases.     

Public-key cryptography is based on \textit{one-way} functions. These are functions whose values are easy to calculate given an argument, but whose inverse is hard to compute. Some such functions enjoy a so-called `trapdoor': supplying an extra piece of information makes the inverse calculation easy. In a public-key system, Bob will create a suitably related pair of a public key and a secret private key. The public key will be used for encryption, which will be easy to perform, but hard to reverse. The private key is the trapdoor that makes the decryption easy. Bob keeps the private key to himself and broadcasts the public key, so that anyone who wants to send him a message may do so, sure in the knowledge that it will be very hard to decrypt by anyone apart from Bob. The best known of such systems is the RSA (Rivest, Shamir and Adlemann) protocol, whose security is based on the apparent computational difficulty of factoring large numbers. The great advantage of public-key systems is that Alice and Bob do not need to meet in secret to share a key---the key used for encryption may simply be broadcast over a public channel. The disadvantage is that the security of the protocol relies only on the computational intractability of the decryption operation in the absence of the private-key; and it's not even known whether any truly adequate one-way functions with trapdoors exist. 

Quantum cryptography, or more properly, \textit{quantum key distribution}, allows one to combine the benefits of both systems. Using quantum systems, Alice and Bob may generate a useable key without having to meet in secret or share any secret beforehand, while at the same time they can be assured of complete security for their communication (at least if the laws of quantum mechanics are correct).

The central idea was first presented by \citet{BB:84}. They realised that one could use the fact that any eavesdropper interacting with quantum systems prepared in non-orthogonal states would disturb those states---and thereby betray their presence---as a basis for sifting out a secret shared random key. The protocol (dubbed `BB84' after its creators) proceeds a follows:
\begin{enumerate}
\item Alice will send Bob a large number of qubits via a quantum channel, choosing at random whether to prepare them in the $\sigma_{z}$ basis or the $\sigma_{x}$ basis (making a note of which she chooses); and choosing at random whether to prepare each system in the up or down spin state (corresponding to a 0 and a 1 value, respectively; again she notes which she chooses). 

\item Bob, on receiving each qubit from Alice, chooses at random whether to measure $\sigma_{z}$ or $\sigma_{x}$ and notes whether he gets a 0 or a 1 (spin-up or spin-down) outcome for each measurement.

Half of the time Bob will have measured in the same basis as Alice prepared the system in; and half of the time he will have measured in a different basis. But neither knows which cases are which. At this stage, both Bob and Alice will possess a random sequence of 0s and 1s, but they will not possess the \textit{same sequence}. If Bob measured in the same basis as Alice chose then the outcome of his measurement will be the same as the value 0 or 1 that Alice prepared, but if he measured in the other basis, he will get a 0 or 1 outcome at random, the value being uncorrelated to the value Alice chose.

\item The next stage of the protocol is that Alice and Bob jointly announce which basis they chose for each system, discarding from their records the bit values for all those systems where they differed in the basis chosen (they do \textit{not}, however, announce their classical bit values). The resulting string of classical bits that Alice and Bob now each possess is called the \textit{sifted key} and, in the absence of noise or any eavesdropping on the transmitted quantum systems, they will now share a secret random key. Notice that neither Alice nor Bob determines which of Alice's initial random sequence of 0 or 1 choices is retained at the sifted key stage; it is a matter of chance depending on the coincidences in their independent random choices of basis.   

\item Now is the time to check for Eve. Given that the qubits sent from Alice to Bob are prepared in a random sequence of states drawn from a non-orthogonal set, any attempt by Eve to determine what the states are will give rise to a disturbance of the sequence. For instance, she might try to gain some information about the key by measuring either $\sigma_{z}$ or $\sigma_{x}$ on each system \textit{en route} between Alice and Bob: this would provide her with some information about the sequence being sent; but half the time it would project the state of a qubit into the other basis than the one Alice initially prepared. Alice and Bob can check for such disturbance by Alice randomly selecting a subset of bits from her sifted key and announcing which bits she has chosen and their values. If the qubits were undisturbed in transmission between Alice and Bob, then Bob should have exactly the same bit values as Alice has announced.

\item Finally, Bob announces whether his bit values for the checked sub-set agree with Alice's or differ. If they agree for the subset of bits publicly announced and checked then Alice and Bob can be sure that there was no eavesdropping; and the remaining bits in their sifted key after they have discarded the checked bits consitute a secret shared random key. If the checked values differ too much, however, then Alice and Bob discard all the remaining bits and recommence the protocol.  

\end{enumerate} 

\noindent Once Alice and Bob have completed the protocol successfully, they know they share a secret random key that can be used for one-time pad encryption. The cryptogram can be broadcast over public channels and Bob (and nobody else) will be able to decrypt it. 

\paragraph{Remarks}
\begin{itemize}
\item[\textbf{a)}] In this protocol, Alice and Bob make use of two channels: a quantum channel transmitting the qubits, which they assume Eve may have access to; and a public (broadcast) channel which anyone can hear, but, we assume, Eve cannot influence. Notice that Eve can always prevent Alice and Bob from successfully completing their protocol and obtaining their key simply by blocking the quantum channel. But this would be self-defeating from her point of view. Her end is to acquire some information about Alice and Bob's random key, so that she may gain some information about any future message they may encrypt using it. If she prevents them from coming to share a key, then they will never try to send such a message, so she would automatically be unable to find out any secrets.    

\item[\textbf{b)}] The crucial component of quantum key distribution is the fact that Eve cannot gain any information about the identity of the states being sent from Alice to Bob without betraying her presence by disturbing them. We saw this in the simple case in which Eve essays an `intercept and resend' strategy: intercepting individual qubits \textit{en route}, measuring them, and then hoping to send on to Bob a new qubit in the same state as the original one sent from Alice, so that her measurement is not detected. In the case where Eve intercepts and measures in either the $\sigma_{z}$ or $\sigma_{x}$ basis, she will introduce $25\%$ errors into the sifted key, which will be easy to detect at the data checking stage ($50\%$ of the systems get projected into the other basis by her measurement; measuring these, half the time Bob will, at random, get a result correlating with Alice's, half the time, however, he will get the opposite result: an error)\footnote{In general, Eve could attempt more subtle attacks, for example, not measuring individual systems, but blocks of them, or entangling ancilla systems with each qubit and not performing any measurement on these ancillas until after Alice and Bob have started making their announcements. Accordingly, full security proofs need to be equally subtle. See, e.g. \citet[\S 12.6.5]{nielsen:chuang} and refs.}.

Notice the links with our previous ideas of no-cloning and of the impossibility of determining an unknown state by measurement (the impossibility of distinguishing perfectly between non-orthogonal states). If Eve were able to clone the qubits sent from Alice to Bob, then she could keep a copy of each for herself and produce her own copy of Alice and Bob's key as they make the crucial announcements; if she could determine unknown states by measurement, she could intercept the qubits, find out what states Alice was sending to Bob and prepare a fresh sequence in the same states afterwards to resend. Whilst it can also be proved directly (see \citet{BBM:92} for a simple case) the fact that Eve must introduce some disturbance when she tries to gain information about the identity of the states being sent can actually be seen as a requirement of \textit{consistency} given the impossibility of distinguishing perfectly between non-orthogonal states (c.f. \citet{busch:observable,fuchs:info-gain}). 

To see why, consider the simple case of a pair of non-orthogonal states $\ket{\phi_{1}}{}$ and $\ket{\phi_{2}}{}$ (the reasoning generalises). A necessary, but not sufficient, condition to be able to distinguish between these states by making some measurement $M$, is that the two states generate different probability distributions over the outcomes of the measurement. We have a system prepared in one or other of these states. Suppose that measuring $M$ did not disturb either $\ket{\phi_{1}}{}$ or $\ket{\phi_{2}}{}$. This would mean that by repeating the measurement over and over again on our individual system, we could eventually arrive at a good estimate of the probability distribution that the state of the system generates, as the state remains the same pre- and post-measurement. But knowing the probability distribution generated for the outcomes of $M$ would allow us to see whether the state of the system was $\ket{\phi_{1}}{}$ or $\ket{\phi_{2}}{}$, given, by hypothesis, that these two distributions are distinct. Thus it \textit{cannot be the case} that neither of these non-orthogonal states is left undisturbed by $M$. Any measurement that would provide information about the identity of the state of the system must therefore lead to a disturbance of at least one of the states in the non-orthogonal set; hence Eve will always betray her presence by introducing errors with some non-zero probability.\footnote{To see how the argument generalises, consider a larger non-orthogonal set $\{\ket{\phi_{i}}{}\}$. Suppose each $\ket{\phi_{i}}{}$ generated a different probability distribution for the outcomes of $M$. Then $M$ must disturb at least one element of $\{\ket{\phi_{i}}{}\}$ and indeed, one element of every pair-wise orthogonal subset of $\{\ket{\phi_{i}}{}\}$. Consider also another measurement $M^{\prime}$ for which at least some of the states of $\{\ket{\phi_{i}}{}\}$, but perhaps not all, generate distinct probability distributions. (This is a minimal condition for a measurement to count as information-gathering for the set.) It's simple to show that the states that generate distinct probability distributions for $M^{\prime}$ cannot all be orthogonal, so there is at least some non-orthogonal pair from $\{\ket{\phi_{i}}{}\}$ that generates distinct distributions for $M^{\prime}$. Applying our previous reasoning, it follows that at least one of these will be disturbed by measurement of $M^{\prime}$.}

\item[\textbf{c)}] Realistic quantum cryptographic protocols have to allow for the possibility of noise. In BB84, errors that are detected at the data checking stage could be due either to Eve, or to noise, or to both. To account for this, \textit{information reconciliation} and \textit{privacy amplification} protocols were developed (see \citet[\S 12.6.2]{nielsen:chuang} and refs therein). Information reconciliation is a process of error correction designed to increase the correlation between Alice's and Bob's strings by making use of the public channel, while giving away as little as possible to Eve. For example, Alice might choose pairs of bits and announce their parity (bit value sum modulo 2), and Bob will announce whether or not he has the same parity for each of his corresponding pairs. If not, they both discard that pair; if they are the same, Alice and Bob both keep the first bit and discard the second. Knowing the parity of the pair won't tell Eve anything about the value of the retained bit. (This example is from \citet{gisin:qcrypt}). After a suitable process of reconciliation, Alice and Bob will share the same key to within acceptable errors, but if some of the original errors were due to Eve, it's possible that she possesses a string which has some correlation to theirs. If the original error rate was low enough, however, Alice and Bob are able to implement privacy amplification, which is a process that systematically reduces the correlation between their strings and Eve's \citep[\S 12.6.2]{nielsen:chuang}.

\end{itemize}

We have focused on one form of quantum key distribution, which proceeds by transmitting qubits prepared in non-orthogonal states. It is also possible to use \textit{entanglement} to generate a key (\citet{ekert:1991}; see also \citet{BBM:92}). Suppose one had a reliable source of entangled systems, for instance a source that could be relied on to generate the spin singlet state \[\ket{\psi^{-}}{}=1/\sqrt{2}(\ket{\up}{}\ket{\down}{}-\ket{\down}{}\ket{\up}{}).\] If a large number of such entangled pairs were produced and one of each pair given to Alice and one to Bob, then Alice and Bob can procede along the same lines as in the BB84 protocol. Each chooses to measure $\sigma_{z}$ or $\sigma_{x}$ at random on each system, obtaining a random sequence of 0 or 1 outcomes. Then just as before, they announce which basis they measured in for each system and discard those outcomes where they did not measure in the same basis, once more obtaining a sifted random key. Again, they may then check for Eve's presence. (In this case, when measuring in the same basis, Bob will get the opposite outcome to Alice's. He can simply perform a bit-flip on every bit to obtain the correlated values.) If they wished to, they could even select a subset of the qubits produced by the source to check that the states being produced by the source violate a Bell inequality---that way they can be sure that sneaky Eve has not replaced the putative singlet source with some other source that might provide her with greater information.      

Quantum key distribution is the aspect of quantum information that has achieved the greatest practical development so far, making use of photon qubits. From the first table-top demonstration models in 1989, key distribution systems have now been demonstrated over distances of tens of kilometers. The DARPA Quantum Network, a quantum key distribution network involving half-a-dozen nodes, has been running continuously since 2004 under the streets of Cambridge Massachussetts, linking Harvard and Boston Universities. Anton Zeilinger's group in Vienna is leading a collaboration (Space-QUEST)involving the European Space Agency, that will see an entangled photon source on the International Space Station by 2012 for the distribution of entanglement to widely separated ground stations from space; a quite remarkable prospect that would allow testing of the properties of entanglement over longer distances than possible on Earth, as well as key distribution between very widely separated sites\footnote{See \texttt{www.quantum.at/quest}.}. 

While quantum cryptography is not exclusively concerned with quantum key distribution, also including discussion of other kinds of protocols such as bit-commitment (of which we will hear a little more later), it is true to say that key distribution has been the dominant interest. It is therefore important to note that in the context of key distribution, quantum cryptography is not concerned with the actual transmission of secret messages, or with hiding messages in quantum systems. Rather, it deals with the problem of establishing certain \textit{necessary conditions} for the classical transmission of secret messages, in a way that could not be achieved classically. The keys that Alice and Bob arrive at after such pains, using their transmitted quantum systems, are not themselves messages, but a means of encoding real messages secretly.

\subsection{Entanglement-assisted communication}

In his lectures Wittgenstein used to say: Don't look for the meaning, look for the use.
Misappropriating gently, we might describe quantum information theorists as adopting just such an attitude vis \`{a} vis entanglement. The strategy has paid-off handsomely. Focusing on what one can \textit{do} with entanglement, considered as a communication and computational resource, the theory of entanglement has blossomed enormously, with the development of a range of quantitative measures of entanglement, intensive study of different kinds of bi-partite and multi-partite entanglement and detailed criteria for the detection and characterisation of entanglement (see \citet{bruss} for a succinct review; \citet{eisertgross:review} for more on multi-particle entanglement).  The conceptual framework provided by questions of communication and computation was essential to presenting the right kinds of questions and the right kinds of tools to drive these developments. 

A state is called entangled if it is not separable, that is, if it cannot be written in the form:
\[ |\Psi\rangle_{AB}=|\phi\rangle_{A}|\psi\rangle_{B}, \textrm{ for pure, or  } \rho_{AB}=\sum_{i}\alpha_{i}\rho_{A}^{i}\otimes \rho_{B}^{i},\textrm{  for mixed states,}\]
where $\alpha_{i} > 0, \sum_{i}\alpha_{i}=1$ and $A$, $B$ label the two distinct subsystems. The case of pure states of bipartite systems is made particularly simple by the existence of the Schmidt decomposition---such states can always be written in the form:
\begin{equation}\label{schmidt}
|\Psi\rangle_{AB}=\sum_{i}\sqrt{p_{i}}\,|\bar{\phi_{i}}\rangle_{A}|\bar{\psi_{i}}\rangle_{B},
\end{equation}
where $\{|\bar{\phi_{i}}\rangle\},\{|\bar{\psi_{i}}\rangle\}$ are orthonormal bases for systems $A$ and $B$ respectively, and $p_{i}$ are the (non-zero) eigenvalues of the reduced density matrix of $A$. The number of coefficients in any decomposition of the form (\ref{schmidt}) is fixed for a given state $|\Psi\rangle_{AB}$, hence if a state is separable (unentangled), there is only one term in the Schmidt decomposition, and conversely. For the mixed state case, this simple test does not exist, but progress has been made in providing operational criteria for entanglement: necessary and sufficient conditions for $2\otimes 2$ and $2\otimes 3$ dimensional systems and necessary conditions for separability (sufficient conditions for entanglement) otherwise \citep{horodeckisPLA:1996,peresseparability}. (See \citet{seevinckuffink:2001, seevincksvetlichny:2002} for discussion of $N$-party criteria.)

\sloppy It is natural to think that shared entanglement could be a useful communication-theoretic resource; that sharing a pair of systems in an entangled state would allow you to do things that you could not otherwise do. (A familiar one: violate a Bell inequality.) The essence of entangled systems, after all, is that they possess global properties that are not reducible to local ones; and we may well be able to utilise these distinctive global properties in trying to achieve some communication task or distributed computational task. The central idea that entanglement---genuinely quantum correlation---differs from any form of classical correlation (and therefore may allow us to do things a shared classical resource would not) is enshrined in the central law (or postulate) of entanglement theory: that the amount of entanglement that two parties share cannot be increased by local operations that each party performs on their own system and classical communication between them. This is a very natural constraint when one reflects that one shouldn't be able to create shared entanglement \textit{ex nihilo}. If Alice and Bob are spatially separated, but share a separable state, then no sequence of actions they might perform locally on their own systems, even chains of conditional measurements (where Bob waits to see what result Alice gets before he choses what he will do; and so on) will turn the separable state into an entangled one. Classical correlations may increase, but the state will remain separable\footnote{If Alice and Bob were in the same location, though, it would be easy for them to turn a separable state into an entangled state, as they can perform operations on the whole of the tensor product Hilbert space (e.g. perform a unitary on the joint space mapping $\ket{\up}{A}\ket{\up}{B}$ to $1/\sqrt{2}(\ket{\up}{A}\ket{\down}{B}-\ket{\down}{A}\ket{\up}{B})$). When spatially separated, they may only perform operations on the individual systems' Hilbert spaces.}. Possessing such a non-classical shared resource, then, we can proceed to ask what one might be able to do with it.  

The two paradigmatic cases of the use of entanglement to assist communication are \textit{superdense coding} \citep{superdense} and \textit{teleportation} \citep{teleportation}.

\subsubsection{Superdense Coding}

Superdense coding is a protocol that allows you to send \textit{classical} information in a surprising way using shared entanglement. If Alice and Bob share a maximally entangled state of two qubits, such as the singlet state, then Alice will be able to transmit to Bob \textit{two} classical bits when she only sends him \textit{one} qubit, twice as much as the maximum we usually expect to be able to send with a single qubit, and apparently in violation of the Holevo bound!

\begin{figure}
\begin{center}\scalebox{0.7}{\includegraphics*{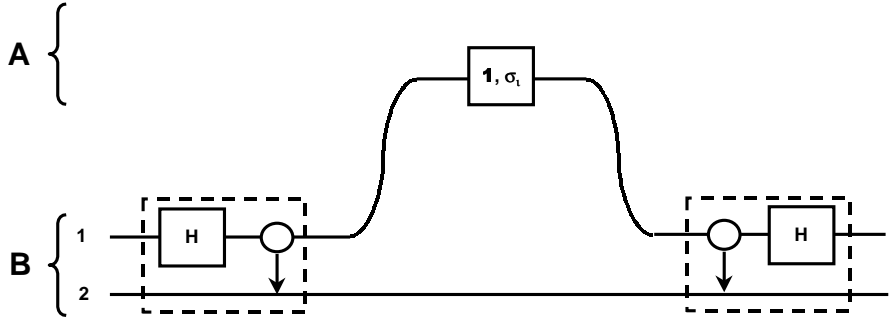}}\end{center}
\caption{\small Superdense coding. Time runs along the horizontal axis. A maximally entangled state of systems 1 and 2 is prepared by Bob (B), here by the action of a Hadamard gate, H, which performs a rotation of $\pi$ around an axis at an angle of $\pi/4$ in the $z$-$x$ plane; followed by a controlled-NOT operation---the circle indicates the control qubit, the point of the arrow, the target, to which $\sigma_{x}$ is applied if the control is in the 0 computational state. System 1 is sent to Alice (A) who may do nothing, or perform one of the Pauli operations. On return of system 1, Bob performs a measurement in the Bell basis, here by applying a controlled-NOT operation, followed by the Hadamard gate. This allows him to infer which operation was performed by Alice.}\label{sdcoding}
\end{figure}

The trick is that Alice may use a local unitary operation to change the global state of the entangled pair. Applying one of the Pauli operators $\{\mathbf{1},\sigma_{x},\sigma_{y},\sigma_{z}\}$ to her half of the entangled pair, she can flip the joint state into one of the others of the four maximally entangled \textit{Bell states} (see Table~\ref{bellstates}), a choice of one from four, corresponding to two bit values (00, 01, 10 or 11). If Alice now sends Bob her half of the entangled pair, he can simply perfom a measurement in the Bell basis to see which of the four states Alice has produced, thereby gaining two bits of information (Fig.~\ref{sdcoding}).

\begin{table}
\begin{center}
\begin{minipage}{2.3in}
\[ \left. \begin{array}{c}
\ket{\phi^{+}}{}=1/\sqrt{2}(\ket{\up}{}\ket{\up}{}+\ket{\down}{}\ket{\down}{})\\
\ket{\phi^{-}}{}=1/\sqrt{2}(\ket{\up}{}\ket{\up}{}-\ket{\down}{}\ket{\down}{})\\
\ket{\psi^{+}}{}=1/\sqrt{2}(\ket{\up}{}\ket{\down}{}+\ket{\down}{}\ket{\up}{})\\
\ket{\psi^{-}}{}=1/\sqrt{2}(\ket{\up}{}\ket{\down}{}-\ket{\down}{}\ket{\up}{})
\end{array} \right\}\]
\end{minipage} \ =  \
\begin{minipage}{1.5in}
\[ \left\{\begin{array}{r}
-i\sigma_{y}\otimes \mathbf{1}\ket{\psi^{-}}{}\\
-\sigma_{x}\otimes \mathbf{1}\ket{\psi^{-}}{}\\
\sigma_{z}\otimes \mathbf{1}\ket{\psi^{-}}{}\\
\mathbf{1}\otimes\mathbf{1}\ket{\psi^{-}}{}
\end{array} \right. \]
\end{minipage}
\end{center}
\caption{The four Bell states, a maximally entangled basis for $2\otimes2$ dim. systems.}\label{bellstates}
\end{table}

But what about the Holevo bound? How can it be that a single qubit is carrying two classical bits in this protocol? The simple answer is that it is not. The presence of \textit{both} qubits is essential for the protocol to work; and it is the pair, as a whole, that carry the two bits of information; therefore there is no genuine conflict with the Holevo bound. What is surprising, perhaps, is the time ordering in the protocol. There would be no puzzle at all if Alice simply encoded two classical bit values into the state of a pair of qubits and sent the pair to Bob (and she could choose any othogonal basis for the pair, whether separable or entangled to do this, so long as Bob knows which she opts for). But although there are two qubits involved in the protocol, Alice doesn't make her choice of classical bit value until one half of the entangled pair is with her and one half with Bob. It then looks puzzling how, when she has access only to one system, she could encode information into both\footnote{The communication in this protocol goes in two steps, first the sharing of the entanglement, then the sending by Alice of her qubit to Bob. One way to think of things is that sharing entanglement is a way of saving up some communication in advance, whose content you can determine later, at any time you wish. Compare the discussion in \citet{mermin:teleportation} of a similar point regarding teleportation.}. And one might think that it \textit{must} be the qubit she sends to Bob that \textit{really} contains the information, from considerations of locality and continuity. 

It turns out that this latter thought rests on a mistake, however, one which also proves significant in understanding teleportation; we will discuss it in Section~\ref{understanding teleportation}. In truth, superdense coding is to be understood in terms of a simple physical mechanism, albeit a non-classical one. The protocol relies on the fact that in the presence of entanglement, local operations can have a non-trivial effect on the global state of the system, that is, can change the irreducibly global properties of the joint system. In particular, it is possible to span bases of maximally entangled states simply by performing local operations \citep{superdense}. Alice, performing her unitary on her system, is able to make a change in the global properties of the joint system; a change, note, that is in fact as great as it could be, flipping the original joint state into one orthogonal to it. It's because of this physical property of maximally entangled states that Alice is able to encode two bit values into the global state of the joint system when she will, and when she only has access to one half of the pair. (See \citet{erpart1} and \citet{nifpaper}  for discussion of whether this sort of phenomenon amounts to a new form of non-locality or not.)  

\subsubsection{Teleportation}

The notion of teleportation is familiar from science fiction: objects are made to disappear (dematerialise) from one location and re-appear (re-materialise) exactly as they were before at another, distant, location. Anyone with a cursory knowledge of quantum mechanics might think that there were fundamental physical reasons why such a process would be impossible. To make something, or someone, re-appear exactly as before, it would seem that we would need to be able to determine their prior physical state exactly. But this would require knowing the quantum states of each individual component of the person or thing, down to the last atom, presumably; and we know that it is just not possible to determine unknown quantum states; and we may well disturb things trying to do so. So teleportation must be physically impossible. But is it? Surprisingly, teleportation does turn out to be possible \textit{if we make use of some entanglement}.   

\begin{figure}
\begin{center}\scalebox{0.7}{\includegraphics*{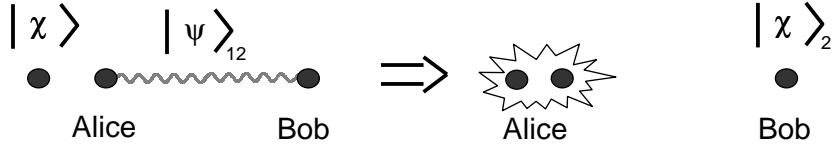}}\end{center}
\caption{\small Teleportation.}\label{minitel}
\end{figure}

In quantum teleportation Alice and Bob again share a pair of particles in a maximally entangled state. If Alice is presented with some system in an unknown quantum state then she is able to make this very state re-appear at Bob's location, while it is destroyed at hers (Fig.~\ref{minitel}). Moreover---and this is the remarkable bit---nothing depending on the identity of the unknown state crosses the region between. Superdense coding uses entanglement to assist classical communication, but in quantum teleportation, entanglement is being used to transmit something purely quantum mechanical---an unknown quantum state, intact, from Alice to Bob. It therefore deserves to be known as the first protocol genuinely concerned with quantum information transmission proper; although we should note that the protocol was devised a little before the full-blown concept of quantum information had been developed by Schumacher. 

Let's consider the standard example using qubits in more detail \citep{teleportation}. We begin with Alice and Bob sharing one of the four Bell states, let's say the singet state $\ket{\psi^{-}}{}$. Alice is presented with a qubit in some unknown state $\ket{\chi}{} = \alpha\ket{\up}{} + \beta\ket{\down}{}$ and her aim is to transmit this state to Bob.

\begin{figure}
\begin{center}\scalebox{0.9}{\includegraphics*{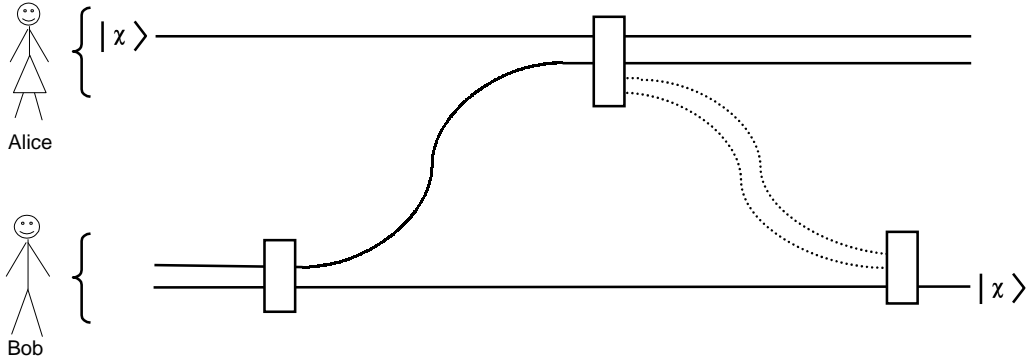}}\end{center}
\caption{\small Teleportation. A pair of systems is first prepared in an entangled state and shared between Alice and Bob, who are widely spatially separated. Alice also possesses a system in an unknown state $\ket{\chi}{}$. Once Alice performs her Bell-basis measurement, two classical bits recording the outcome are sent to Bob, who may then perform the required conditional operation to obtain a system in the unknown state $\ket{\chi}{}$. (Continuous black lines represent qubits, dotted lines represent classical bits.)\label{telep1}}
\end{figure}

By performing a suitable joint measurement on her half of the entangled pair and the system whose state she is trying to transmit (in this example, a measurement in the Bell basis), Alice will change the state of Bob's half of the entangled pair into a state that differs from $\ket{\chi}{}$ by one of four unitary transformations, depending on what the outcome of her measurement was. If a record of the outcome of Alice's measurement is then sent to Bob, he may perform the required operation to obtain a system in the state Alice was trying to send (Fig.~{\ref{telep1}}). 

The end result of the protocol is that Bob obtains a system in the state $\ket{\chi}{}$, with nothing that bears any relation to the identity of this state having traversed the space between him and Alice. Only two classical bits recording the outcome of Alice's measurement were sent between them; and the values of these bits are completely random, with no dependence on the parameters $\alpha$ and $\beta$. Meanwhile, no trace of the identity of the unknown state remains in Alice's region, as required, of course, to accord with the no-cloning theorem (the state of her original system will usually now be maximally mixed). The state has indeed disappeared from Alice's region and reappeared in Bob's, so `teleportation' really does seem an appropriate name for this phenomenon.

The formal description of the process is straightforward. We begin with system 1 in the unknown state $\ket{\chi}{}$ and Alice and Bob sharing a pair of systems (2 and 3) in the singlet state $\ket{\psi^{-}}{}$. The total state of the three systems at the beginning of the protocol is therefore simply
\begin{equation}\label{start}
\ket{\chi}{1}\ket{\psi^{-}}{23} = \frac{1}{\sqrt 2} \bigl(\alpha \ket{\up}{1} + \beta\ket{\down}{1}\bigr)\bigl(\ket{\up}{2}\ket{\down}{3}-\ket{\down}{2}\ket{\up}{3}\bigr).
\end{equation} 
Notice that at this stage, the state of system 1 factorises from that of systems 2 and 3; and so in particular, the state of Bob's system is independent of $\alpha$ and $\beta$. We may re-write this initial state in a suggestive manner, though:

\begin{singlespacing}
\begin{align}
\ket{\chi}{1}\ket{\psi^{-}}{23} & = \frac{1}{\sqrt{2}}\biggl(\alpha\ket{\up}{1}\ket{\up}{2}\ket{\down}{3} + \beta\ket{\down}{1}\ket{\up}{2}\ket{\down}{3} - \alpha\ket{\up}{1}\ket{\down}{2}\ket{\up}{3} - \beta\ket{\down}{1}\ket{\down}{2}\ket{\up}{3}\biggr) \\
   \begin{split} & =  
 \frac{1}{2}\biggl(\ket{\phi^{+}}{12} \bigl(\alpha \ket{\down}{3} - \beta\ket{\up}{3}\bigr)+\ket{\phi^{-}}{12} \bigl(\alpha \ket{\down}{3} + \beta\ket{\up}{3}\bigr) \\
 & \phantom{\ket{\phi^{+}}{} }+ \ket{\psi^{+}}{12} \bigl(-\alpha \ket{\up}{3} + \beta\ket{\down}{3}\bigr) + \ket{\psi^{-}}{12} \bigl(-\alpha \ket{\up}{3} - \beta\ket{\down}{3}\bigr)\biggr). 
\end{split} \label{rewrite1}
\end{align}\end{singlespacing}
The basis used is the set \[\{\ket{\phi^{\pm}}{12}\ket{\up}{3},\,\ket{\phi^{\pm}}{12}\ket{\down}{3},\, \ket{\psi^{\pm}}{12}\ket{\up}{3}, \, \ket{\psi^{\pm}}{12}\ket{\down}{3}\},\] that is, 
we have chosen (as we may) to express the total state of systems 1,2 and 3 using an entangled basis for systems 1 and 2, even though these systems are quite independent. But so far, of course, all we have done is re-written the state in a particular way; nothing has changed physically and it is still the case that it is really systems 2 and 3 that are entangled and wholly independent of system 1, in its unknown state.

Looking closely at (\ref{rewrite1}) we notice that the relative states of system 3 with respect to particular Bell basis states for 1 and 2 have a very simple relation to the initial unknown state $\ket{\chi}{}$; they differ from $\ket{\chi}{}$ by one of four local unitary operations:
\begin{multline}\label{rewrite2}
\ket{\chi}{1}\ket{\psi^{-}}{23} = \frac{1}{2}\biggl( \ket{\phi^{+}}{12} \bigl(-i\sigma_{y}^{3}\ket{\chi}{3}\bigr) + \ket{\phi^{-}}{12} \bigl(\sigma_{x}^{3}\ket{\chi}{3}\bigr) \\
 + \ket{\psi^{+}}{12} \bigl(-\sigma_{z}^{3}\ket{\chi}{3}\bigr) + \ket{\psi^{-}}{12} \bigl(-\mathbf{1}^{3}\ket{\chi}{3}\bigr)\biggr),  
\end{multline}
where the $\sigma_{i}^{3}$ are the Pauli operators acting on system 3 and $\mathbf{1}$ is the identity. To re-iterate, though, only system 1 actually depends on $\alpha$ and $\beta$; the state of system 3 at this stage of the protocol (its reduced state, as it is a member of an entangled pair) is simply the maximally mixed $1/2\,\mathbf{1}$.  

Alice is now going to perform a measurement. If she were simply to measure system 1 then nothing of interest would happen---she would obtain some result and affect the state of system 1, but systems 2 and 3 would remain in the same old state $\ket{\psi^{-}}{}$. However, as she has access to both systems 1 and 2, she may instead perform a \textit{joint} measurement, and now things get interesting. In particular, if she measures 1 and 2 in the Bell basis, then after the measurement we will be left with only one of the terms on the right-hand side of eqn.~(\ref{rewrite2}), at random; and this means that Bob's system will have jumped instantaneously into one of the states $-i\sigma_{y}^{3}\ket{\chi}{3},\, \sigma_{x}^{3}\ket{\chi}{3},\, -\sigma_{z}^{3}\ket{\chi}{3}$ or  $-\ket{\chi}{3}$, with equal probability.

But how do things look to Bob? As he neither knows whether Alice has performed her measurement, nor, if she has, what the outcome turned out to be, he will still ascribe the same, original, density operator to his system---the maximally mixed state\footnote{Notice that an equal mixture of the four possible post-measurement states of his system results in the density operator $1/2\, \mathbf{1}$.}. No measurement on his system could yet reveal any dependence on $\alpha$ and $\beta$. To complete the protocol therefore, Alice needs to send Bob a message instructing him which of four unitary operators to apply $(i\sigma_{y},\,\sigma_{x},\,-\sigma_{z},\,\mathbf{-1})$ in order to make his system acquire the state $\ket{\chi}{}$ with certainty; for this she will need to send two bits\footnote{Two bits are clearly sufficient, for the argument that they are strictly necessary, see \citet{teleportation} Fig.2.}. With these bits in hand, Bob applies the needed transformation and obtains a system in the state $\ket{\chi}{}$.\footnote{In this description, as in the the original \citet{teleportation} treatment, we have assumed that a process of collapse occurs after Alice's measurement, in order to pick out, probabilistically, a definite state of Bob's system. It is straightforward, however, to give no-collapse versions of the teleportation protocol. \citet{vaidman} provides an Everettian description and \citet{braunstein:irreversible} a detailed general discussion of teleportation in a no-collapse setting. See \citet{grammar} for further discussion.}

We should note that this quantum mechanical process differs from science fiction versions of teleportation in at least two ways, though. First, it is not \textit{matter} that is transported, but simply the quantum state $\ket{\chi}{}$; and second, the protocol is not instantaneous, but must attend for its completion on the arrival of the classical bits sent from Alice to Bob. Whether or not the quantum protocol approximates to the science fiction ideal, however, it remains a very remarkable phenomenon from the information-theoretic point of view\footnote{Interestingly, it can be argued that quantum teleporation is perhaps not so far from the sci-fi ideal as one might initially think. \citet{vaidman} suggests that if all physical objects are made from elementary particles, then what is distinctive about them is their form (i.e. their particular state) rather than the matter from which they are made. Thus it seems one could argue that \textit{objects} really are teleported in the protocol.}. For consider what has been achieved. An unknown quantum state has been sent to Bob; and how else could this have been done? Only by Alice sending a quantum system \textit{in} the state $\ket{\chi}{}$ to Bob\footnote{Or by her sending Bob a system in a state explicitly related to $\ket{\chi}{}$ (cf. \citet{park:1970}).}, for she cannot determine the state of the system and send a description of it instead. 

If, however, Alice did \textit{per impossibile} somehow learn the state and send a description to Bob, then systems encoding that description would have to be sent between them. In this case something that \textit{does} bear a relation to the identity of the state is transmitted from Alice to Bob, unlike in teleportation. Moreover, sending such a description would require a \textit{very great deal} of classical information, as in order to specify a general state of a two dimensional quantum system, two \textit{continuous} parameters need to be specified.

The picture we are left with, then, is that in teleportation there has been a transmission of something that is inaccessible at the classical level; in the transmission this information has been in some sense disembodied; and finally, the transmission has been very efficient---requiring, apart from prior shared entanglement, the transfer of only two classical bits. The initial entanglement that Alice and Bob shared, however, will have been used up at the end of the protocol. If Alice wanted to teleport any more unknown states to Bob, they would need to be in possession of more entangled pairs.

While the formal description of teleportation is, as we have seen, simple, the question of how one ought to understand what is going on has been extremely vexed. We will return to this question in Section~\ref{understanding teleportation}. It is worth noting, however, that teleportation, just like superdense coding, is driven by the fact that local operations can induce substantive differences in global properties of entangled systems \citep{braunstein:twist}; again, specifically, by the fact that maximally entangled bases can be spanned by local unitary operations.

Finally, we should note that since teleportation is a linear process, it may be used for the process of \textit{entanglement swapping}. Let's suppose that Alice shares one maximally entangled state with Bob and another with Charles. If she performs the teleportation protocol on her half of the Alice-Charles entangled pair, then the result will be that the initial entanglement between Alice and Bob will be destroyed, and the initial entanglement between Alice and Charles will be destroyed, but Charles and Bob will now share a maximally entangled pair when they did not before. Thus entanglement can be swapped from Alice-Charles to Charles-Bob, at the cost of using up an entangled pair that Alice and Bob shared.

\subsubsection{Quantifying entanglement}

The basic examples we have seen of superdense coding and teleportation both make use of maximally entangled pairs of qubits. If the qubits were less than maximally entangled then the protocols would not work properly, perhaps not at all. Given that entanglement is a communication resource that will be used up in a process like teleportation, it is natural to want to quantify it. The amount of entanglement in a Bell state, the amount required to perform teleportation of a qubit, is defined as one \textit{ebit}. The general theory of quantifying entanglement takes as its central axiom the condition that we have already met: no increase of entanglement under local operations and classical communication. In the case of pure bipartite entanglement, the measure of degree of entanglement turns out to be effectively unique, given by the von Neumann entropy of the reduced states of the entangled pair \citep{popescurohrlich:uniqueness,donald:uniqueness}. In the case of mixed state entanglement, there exists a range of distinct measures. \citet{vlatko:quantifying,vlatko:measures} propose criteria that any adequate measure must satisfy and discuss relations between a number of measures.

\subsection{Quantum computers}

Richard Feynman was the prophet of quantum computation. He pointed out that it seems that one cannot simulate the evolution of a quantum mechanical system efficiently on a classical computer. He took this to imply that there might be computational benefits to be gained if computations are carried out using quantum systems themselves rather than classical systems; and he went on to describe a universal quantum simulator \citep{feynman:1982}. However it is with Deutsch's introduction of the concept of the universal qauntum computer that the field really begins \citep{deutsch:1985}.

In a quantum computer, we want to use quantum systems and their evolution to perform computational tasks. We can think of the basic components of a quantum computer as a register of qubits and a system of computational gates that can be applied to these qubits to perform various evolutions and evaluate various functions. States of the whole register of qubits in the computational basis would be $\ket{0}{}\ket{0}{}\ket{0}{}\ldots\ket{0}{}$, for example, or $\ket{0}{}\ket{1}{}\ket{0}{}\ldots\ket{1}{}$, which can also be written $\ket{000\ldots 0}{}$ and $\ket{010\ldots 1}{}$ respectively; these states are analogous to the states of a classical register of bits in a normal computer. At the end of a computation, one will want the register to be left in one of the computational basis states so that the result may be read out.

The immediately exciting thing about basing one's computer on qubits is that it looks as if they might be able to provide one with massive parallel processing. Suppose we prepared each of the $N$ qubits in our register in an equal superposition of 0 and 1, then the state of the whole register will end up being in an equal superposition of all the $2^{N}$ possible sequences of 0s and 1s:
\[\frac{1}{\sqrt{2^{N}}}(\ket{0000\ldots 00}{} + \ket{0000\ldots 01}{} + \ket{0000\ldots 11}{} + \ldots + \ket{1111\ldots 11}{}).\] 

A classical $N$-bit register can store one of $2^{N}$ numbers: an $N$-qubit register looks like it might store $2^{N}$ numbers simultaneously, an enormous advantage. Now if we have an operation that evaluates a function of an input string, the linearity of quantum mechanics ensures that if we perform this operation on our superposed register, we will evaluate the function simultaneously for all possible inputs, ending up with a register in which all the $2^{N}$ outputs are superposed!

This might look promising, but the trouble is, of course, that it is not possible to read out all the values that are superposed in this state. Measuring in the computational basis to read out an outcome we will get a ``collapse" to some one of the answers, at random. Thus despite all the quantum parallel processing that went on, it proves very difficult to read much of it out. In this naive example, we have done no better than if we had evaluated the function on a single input, as classically. It is for this reason that the design of good quantum algorithms is a very difficult task: one needs to make subtle use of other quantum effects such as the constructive and destructive interference between different computational paths in order to make sure that we can read out useful information at the end of the computation, i.e., that we can improve on the efforts of classical computers.

The possible evolutions of states of quantum mechanical systems are given by unitary operators. A \textit{universal} quantum computer will thus be a system that can (using finite means) apply any unitary operation to its register of qubits. It turns out that a relatively small set of one and two qubit \textit{quantum gates} is sufficient for a universal quantum computer\footnote{See for example \citet[\S 4.5]{nielsen:chuang}. We are considering the \textit{quantum network} model of quantum computation which is more intuitive and more closely linked to experimental applications than the alternative \textit{quantum Turing machine} model that Deutsch began with. The two models were shown to be equivalent in \citet{yao:1993}.}. A quantum gate is a device that implements a unitary operation that acts on one or more qubits (we have already seen some schematic examples in Figs.~\ref{sdcoding}~and~\ref{telep1}). By combining different sequences of gates (analogously to logic gates in a circuit diagram) we can implement different unitary operations on the qubits they act on. A set of gates is \textit{universal} if by combining elements of the set, we can build up any unitary operation on $N$ qubits to arbitrary accuracy.

So what can quantum computers do? First of all, they can compute anything that a classical Turing machine can compute; such computations correspond to permutations of computational basis states and can be achieved by a suitable subset of unitary operations. Second, they can't compute anything that a classical Turing machine can't. This is most easily seen in the following way \citep{ekertjozsa:1996}.

We can picture a probabilistic Turing machine as following one branch of a tree-like structure of computational paths, with the nodes of the tree corresponding to computational states. The edges leading from the nodes correspond to the different computational steps that could be made from that state. Each path is labelled with its probability and the probability of a final, halting, state is given by summing the probabilities of each of the paths leading to that state. We may see a quantum computer in a similar fashion, but this time with the edges connecting the nodes being labelled with the appropriate probability amplitude for the transition. The quantum computer follows all of the different computational paths at once, in a superposition; and because we have probability \textit{amplitudes}, the possibility of interference between the different computational paths exists. However, if we wished, we could program a classical computer to calculate the list of configurations of the quantum computer and calculate the complex numbers of the probability amplitudes. This would allow us to calculate the correct probabilities for the final states, which we could then simulate by tossing coins. Thus a quantum computer could be simulated by a probabilistic Turing machine; but such a simulation is very inefficient.

The advantage of quantum computers lies not, then, with what can be computed, but with its efficiency. In computational complexity, the crudest measure of whether a computational task is tractable or not, or an algorithm efficient, is given by seeing how the resources required for the computation scale with increased input size. If the resources scale polynomially with the size of the input in bits, the task is deemed tractable. If they do not, in which case the resources are said to depend exponentially on the input size, the task is called hard or intractable. A breakthrough in quantum computation was achieved when \citet{shor} presented an efficient algorithm for factoring on a quantum computer, a task for which it is believed no efficient classical algorithm exists\footnote{Thus quantum computers would destroy the security of the widely-used RSA public-key protocol mentioned earlier. It's therefore perhaps comforting that what quantum mechanics takes with one hand (ease of factoring, therefore violating state-of-the-art security) it gives back with the other (quantum key distribution).}. Hence quantum computers provide exponential speed-up over the best known classical algorithms for factoring; and this is strong evidence that quantum computers are more powerful than classical computers. Another very important quantum algorithm is due to \citet{grover}. This algorithm also provides a speed-up, although not an exponential one, over classical methods for searching an unstructured database. For a database of size $n$, the algorithm allows the desired object to be found in $\sqrt{n}$ steps, rather than the order of $n$ steps one would expect classically. (A good review of quantum computation up to and including the development of Shor's algorithm is provided by \citet{ekertjozsa:1996}.)

\section{The concept(s) of information}\label{info}

Having reviewed some of the basic features of quantum information theory it's time we were a little more precise about some conceptual matters, specifically, more precise about just what information in this theory is supposed to be. `Information' is a notoriously promiscuous term with a marked capacity for dulling critical capacities: it is used in different ways in a large variety of different contexts across the sciences and in everyday life, in various technical and non-technical uses; and typically little more than lip service is paid to the ensuing conceptual distinctness of these various uses. Often the introduction of a neologism would be preferable to taxing further the sadly over-worked `information'. 

Here we will concern ourselves with the question: What is quantum information? It is commonly supposed that this question has not yet received, perhaps cannot be expected to receive, a definite or illuminating answer. \textit{Vide} the Horodeckis:
\begin{quote} Quantum information, \textit{though not precisely defined}, is a fundamental concept of quantum information theory. \citep{horodeckis:peresdedic}
\end{quote}  
And Jozsa:
\begin{quote}\sloppypar
$\ket{\psi}{}$ may be viewed as a carrier of ``quantum information" which...we leave...undefined in more fundamental terms...Quantum information is a new concept with no classical analogue...In more formal terms, we would aim to formulate and interpret quantum physics in a way that has a concept of information as a primary fundamental ingredient. Primary fundamental concepts are \textit{ipso facto} undefined (as a definition amounts to a characterization in yet more fundamental terms) and they acquire meaning only afterward, from the structure of the theory they support. \citep{jozsa:2003}
\end{quote}
However, I shall demure from this. Given a proper understanding of the meaning and significance of the coding theorems, it becomes clear that quantum information already admits of a perfectly precise and adequate definition; and moreover, that there exist very strong analogies (\textit{pace} Jozsa) between classical and quantum information. Both may be seen as species of a single genus. In addition, the ontological status of quantum information can be settled: I shall argue that quantum information is not part of the material contents of the world. In both classical and quantum information theory, we will see, the term `information' functions as an abstract, not a concrete, noun\footnote{This lesson already features in related ways in \citet{thesis,nifpaper,grammar}.}.
 
\subsection{Coding theorems: Both \textit{what} and \textit{how much}}\label{whathow}

Discussions of information theory, quantum and classical, generally begin with an important caveat concerning the scope of their subject matter. The warnings typically take something like the following form:
\begin{quote}
Note well, reader: Information theory doesn't deal with the \textit{content} or \textit{usefulness} of information, rather it deals only with the \textit{quantity} of information.
\end{quote}
Now while there is obviously an important element of truth in statements such as these, they can also be seriously misleading, in two interrelated ways. First, the distinction between the technical notions of information deriving from information theory and the everyday semantic/epistemic concept is not sufficiently noted; for it may easily sound as if information theory does at least describe the \textit{amount} of information in a semantic/epistemic sense that may be around. But this is not so. In truth we have two quite distinct concepts (or families of concepts)---call them `information$_{e}$' and `information$_{t}$' for the everyday and technical concepts respectively---and quantifying the amount of the latter does not tell us about the quantity, if any, of the former, as Shannon himself noted \citep[p.31]{shannon}. For elaboration on the distinctness of information$_{e}$ and information$_{t}$, including discussion of the opposing view of \citet{dretske:1981}, see \citet[chpt. 1]{thesis}.

The second point of concern is that the coding theorems that introduced the classical \citep{shannon} and quantum \citep{qcoding} concepts of information$_{t}$ do not \textit{merely} define measures of these quantities. They also introduce the concept of \textit{what it is} that is transmitted, \textit{what it is} that is measured. Thus we may as happily describe what information$_{t}$ is, as how much of it there may be. 
Let us proceed to do so.

We may take our lead from Shannon:
\begin{quote}
The fundamental problem of communication is that of reproducing at one point either exactly or approximately a message selected at another. \citep[p.31]{shannon}
\end{quote}
The technical notion of information then enters when we note that information$_{t}$ may be identified as what it is the aim of such a communication protocol to transmit. Thus the following definition suggests itself \citep[\S 1.2.3]{thesis}:
\begin{quote}
Information$_{t}$ is what is produced by an information$_{t}$ source that is required to be reproducible at the destination if the transmission is to be counted a success.
\end{quote} 

This definition is evidently a very general one, but that is as it should be. If we follow Shannon in his specification of what the problem of communication \textit{is}, then the associated notion of information$_{t}$ introduced should be sensitive to what one's aims and interests in setting up a communication system are. Different aims and interests may give rise to more or less subtly differentiated concepts of information$_{t}$ as what one is interested in transmitting and reproducing varies: indeed we will see the most vivid example of this when comparing classical and quantum information$_t$. Yet these all remain concepts of \textit{information}$_{t}$ as they all arise in the general setting adumbrated by Shannon that the broad definition seeks to capture. 

There are several components to the generality of this definition. One might ask what information$_{t}$ sources are; what they produce; and what counts as success. The answers given to these questions, though, will in general be interdependent (we will see some examples below). What counts as a successful transmission will, of course, depend once more upon what one's aims and interests in devising the communication protocol are. Specifying what counts as success will play a large part in determining what it is we are trying to transmit; and this, in turn, will determine what it is that information$_{t}$ sources produce that is the object of our interest. Finally, information$_{t}$ sources will need to be the sorts of things that produce what it is that we are concerned to transmit.

\subsubsection{Two types of information$_{t}$ source}

Some examples will help put flesh on the remarks so far. The prototypical information$_{t}$ source was introduced by Shannon in his noiseless coding theorem. Such a source is some object which may be characterised as producing elements drawn from a fixed alphabet, say a discrete alphabet $\{a_{1},a_{2},\ldots, a_{n}\}$, with given probabilities $p(a_{i})$. (The extension to the continuous case takes the obvious form.) \textit{Messages} are then long sequences of elements drawn from the alphabet. The aim of the communication protocol is to be able to reproduce at some distant point whatever sequence the source produces. 

If classical information$_{t}$ is what is produced by a classical information$_{t}$ source---the Shannon prototype---then quantum information$_{t}$ is what is produced by a \textit{quantum} information$_{t}$ source. Schumacher's notion of a quantum information$_{t}$ source is the immediate generalisation to the quantum domain of the Shannon prototype: A quantum information$_{t}$ source is some object which may be characterised as producing systems in quantum states drawn from a fixed set of states, e.g., $\{\rho_{a_{1}},\rho_{a_{2}},\ldots, \rho_{a_{n}}\}$, with probabilities $p(a_{i})$. Again, we will be interested in long sequences drawn from the source.

We are now in a position to give a general answer to the question of what information$_{t}$ sources produce: they produce sequences of states. Or more precisely, they produce \textit{tokens} of particular \textit{types}.

\paragraph{Classical information$_{t}$}

Let us look more closely at the example of classical information$_{t}$. As we know, a distinguishing characteristic of classical information$_{t}$ when compared with quantum information$_{t}$ is that the varying outputs of a classical information source are distinguishable one from another, i.e., one can tell which of the possible elements $a_{i}$ was produced in a given instance. After the source has run for a while, a given sequence of states will have been produced, for example a sequence like:\[a_{7}a_{3}a_{4}a_{9}a_{9}a_{7}a_{1}\ldots a_{2}a_{1}a_{3}a_{7}\ldots a_{1}a_{9}a_{1}.\]
This particular sequence could be identified by description (e.g., ``It's the sequence `$a_{7}a_{3}a_{4}a_{9}\ldots$'," etc.), by name (call it `sequence 723'), or, given the distinguishability of the $a_{i}$, identified demonstratively. (Handed a concrete token of the sequence, one could in principle determine---generally, infer---what particular sequence it was.) 

This sequence (type) will have been \textit{realised} by a given system, or systems, taking on the properties that correspond to being in the various states $a_{i}$, in order. What will be required at the end of the communication protocol is either that another token of this type actually be reproduced at a distant point; or at least, that it be \textit{possible} to reproduce it there, by a standard procedure.

But what is the information$_{t}$ produced by the source that we desire to transmit? Is it the sequence type, or the token? The answer is quick: it is the type; and we may see why when we reflect on what it would be to specify what is produced and what is transmitted. We would specify what is produced (transmitted) by naming or otherwise identifying the sequence itself---it was sequence 723, the sequence `$a_{7}a_{3}a_{4}a_{9}\ldots$', in the example---and this is to identify the type, not to identify or name a particular concrete instance of it\footnote{Even when we identify what was produced by gesturing to the concrete token and saying `That was what was produced', we are identifying the sequence type, here by means of what Quine would call `deferred ostension'. The `what' in these contexts is functioning as an interrogative, not a relative, pronoun (c.f. \citet[p.76]{glock:qd} for an analogous case).}. 

\paragraph{Quantum information$_{t}$}
The quantum example is similar, but here we must distinguish two cases.

The basic type of quantum information$_{t}$ source \citep{qcoding} is one which produces \textit{pure} states: we may take as our example a device which outputs systems in one of the states $\{\ket{a_{1}}{},\ket{a_{2}}{},\ldots, \ket{a_{n}}{}\}$ with probabilities $p(a_{i})$; these states need not be orthogonal. Then the output of this source after it has been running for a while might be a sequence of systems in particular \textit{quantum} states, e.g.,
\[\ket{a_{7}}{}\ket{a_{3}}{}\ket{a_{4}}{}\ket{a_{9}}{}\ket{a_{9}}{}\ket{a_{7}}{}\ket{a_{1}}{}\ldots \ket{a_{2}}{}\ket{a_{1}}{}\ket{a_{3}}{}\ket{a_{7}}{}\ldots \ket{a_{1}}{}\ket{a_{9}}{}\ket{a_{1}}{}.\]
Again we have a sequence type, instantiated by particular systems taking on various states. And again such a sequence may be named or described, but notice that this time it will not, in general, be possible to identify \textit{what} sequence a given number of systems instantiate merely by being presented with them, as the $\ket{a_{i}}{}$ need not be orthogonal, so typically will not be distinguishable. However, this does not stop the lesson learnt above applying once more: the information$_{t}$ produced by the source---quantum information$_{t}$, now---will be specified by specifying what sequence (type) was produced. These sequences will clearly be of a different, and more interesting, sort than those produced by a classical source. (One might say that with classical and quantum information$_{t}$, one was concerned with different types of type!) Just as before, though, what will be required for a successful transmission to be effected is that another token of this type be reproduced, or be reproducible (following a standard procedure) at the desired destination. That is, we need to be able to end up with a sequence of systems taking on the appropriate quantum states in the right order. What is transmitted is a particular sequence of quantum states.

This was the most basic form of quantum information$_{t}$ source. We gain a richer notion when we take into account the possibility of \textit{entanglement}. So consider a different type of quantum information$_{t}$ source \citep{qcoding}, one that always outputs systems in a particular mixed state $\rho$. Such a source might seem dull until we reflect that these might be systems in \textit{improperly} mixed states \citep{d'Espagnat}, that is, components of larger entangled systems, the other parts of which may be inaccessible to us. In particular, there could be a variety of \textit{different} states of these larger systems that give rise to the same reduced state for the smaller components that the information$_{t}$ source presents us with. How should we conceive of what \textit{this} information$_{t}$ source produces?

We have a choice. We might be unimaginative and simply require that the `visible' output of the source be reproducible at the destination. The source produces a sequence $\rho\otimes\rho\otimes\rho\otimes\ldots$ and we should be able to reproduce this sequence at the destination. What is transmitted will then be specified by specifying this sequence. But we might be more interesting and require that not only should the `visible' output sequence be reproducible at the destination, but so also should any entanglement that the original output systems might possess. Given the importance of being able to transfer entanglement in much of quantum information$_{t}$ theory, this latter choice turns out to be the better one to make\footnote{As \citet{duwell:qinfoexists} has emphasised, this corresponds to the choice of the \textit{entanglement fidelity} (c.f. \citet[Section 9.3]{nielsen:chuang}) as the criterion of successful message reproduction for quantum information$_{t}$.}.

We may model the situation as follows. Take three sets of systems, labelled $A$, $B$ and $C$. Systems in set $B$ are the systems that our source outputs, we suppose them all to be in the mixed state $\rho$. Systems in set $A$ are the hidden partners of systems in set $B$. The $i$th member of $B$ ($B_{i}$) can be thought to be part of a larger system whose other part consists of the $i$th member of $A$ ($A_{i}$); in addition, we assume that the joint system composed of $A_{i}$ and $B_{i}$ together is in some pure state $\ket{\psi}{A_{i}B_{i}}$ which will give a reduced state of $\rho$ when we trace over $A_{i}$ (such a state is called a \textit{purification} of $\rho$). If $\rho$ is mixed then $\ket{\psi}{A_{i}B_{i}}$, by assumption pure, will necessarily be entangled. The systems in set $C$ are the `target' systems at the destination point.

Now consider the $i$th output of our information$_{t}$ source. This will be the system $B_{i}$, having the reduced state $\rho$. But this is only half the story: along with $B_{i}$ is the hidden system $A_{i}$; and together these are in the state $\ket{\psi}{A_{i}B_{i}}$. As the end result of the transmission process, we would like $C_{i}$ to be in the state $\rho$, but if we are to preserve entanglement, then our truly desired end result would be $C_{i}$ becoming entangled to $A_{i}$, in just the way $B_{i}$ had been previously. So we actually desire that the pure state $\ket{\psi}{}$ previously instantiated by $A_{i}B_{i}$ should end up being instantiated by $A_{i}$ and $C_{i}$. This would be transfer of the entanglement, or transfer of the `quantum correlation', that $B_{i}$---the visible output of the source---had previously possessed.

This may all now be expressed in terms of sequences of states once more. The quantum source outputs sequences of systems in entangled states, half of which (systems $B$) we see; and half of which (systems $A$) we do not. A particular segment of such a sequence might look like:
\[ \ldots \ket{\psi}{A_{i}B_{i}}\ket{\psi^{\prime}}{A_{j}B_{j}}\ket{\psi^{\prime\prime}}{A_{k}B_{k}}\ldots,\]
where $\ket{\psi^{\prime}}{}$ and $\ket{\psi^{\prime\prime}}{}$, like $\ket{\psi}{}$, are purifications of $\rho$. Such a sequence is the piece of quantum information$_{t}$ produced and it will be successfully reproduced by a protocol if the end result is another token of the type, but this time involving the systems $C$:
\[ \ldots \ket{\psi}{A_{i}C_{i}}\ket{\psi^{\prime}}{A_{j}C_{j}}\ket{\psi^{\prime\prime}}{A_{k}C_{k}}\ldots.\]

The general conclusion we may draw is that pieces of quantum information$_{t}$, far from being mysterious---perhaps unspeakable---are quite easily and perspicuously described. A given item of quantum information$_{t}$ will simply be some particular sequence of Hilbert space states, whether the source produces systems in individual pure states, or as parts of larger entangled systems. What is more, we have seen that quantum information$_{t}$ is closely analogous to classical information$_{t}$: in both cases, information$_{t}$ is what is produced by the respective information$_{t}$ sources (both fall under the general definition); and in both cases, what is produced can be analysed in terms of sequences of states (types).

\subsection{Bits and pieces}
So far we have been emphasising the largely neglected point that the coding theorems characteristic of information$_{t}$ theory provide us with a perfectly good and straightforward account of what information$_{t}$ is; but we should not, in our enthusiasm, forget the more commonly emphasised aspect of these theorems. It is also of the utmost importance that the coding theorems provide us with a notion of \textit{how much} information$_{t}$ a given source outputs. How much information$_{t}$ a source produces is measured, following Shannon, in terms of the minimal amount of channel resources required to encode the output of the source in such a way that any message produced may be accurately reproduced at the destination. That is, to ask how much information$_{t}$ a source produces is ask to what degree is the output of the source \textit{compressible}? Shannon showed that the compressibility of a classical information$_{t}$ source is given by the familiar expression
\[H(A) = -\sum_{i} p(a_{i})\log p(a_{i}),\]
known as the Shannon information$_{t}$ (logarithms to base 2). This specifies the number of \textit{bits} required per letter to encode the output of the source. \citet{qcoding} extended this proof to the quantum domain, showing that the minimum number of \textit{qubits} required per step to encode the output of quantum information$_{t}$ sources of the sorts mentioned above, is given by the von Neumann entropy of the source:
\[S(\rho) = -\textrm{Tr}\rho\log\rho,\]
where $\rho$ is the density matrix associated with the output of the source.

So this aspect of the coding theorems provides us with the notion of \textit{bits} of information, quantum or classical; the amount of information$_{t}$ that a source produces; and this is to be contrasted with \textit{pieces} of information$_{t}$, \textit{what} the output of a source (quantum or classical) is, as described above.

\subsection{The worldliness of quantum information}

Let us now consider an important corollary of the discussion so far. It concerns the worldliness or otherwise of information$_{t}$. Is information$_{t}$ part of the material contents of the world? In particular, is \textit{quantum} information$_{t}$ part of the material contents of the world? Is it a new type of physical substance or stuff, admittedly, perhaps, a rather unusual one, that has a spatio-temporal location and whose ebb and flow it is the aim of quantum information$_{t}$ theory to describe? The writings of some physicists \citep[for example]{jozsa:1998, penrose:1998, dh} might lead one to suppose so. However it follows from the analysis in Section~\ref{whathow} that this thought would be mistaken.

Information$_{t}$, what is produced by a source, or what is transmitted, is not a concrete thing or a stuff. It is not so, because, as we have seen, what is produced/transmitted is a sequence \textit{type} and types are \textit{abstracta}. They are not themselves part of the contents of the material world, nor do they have a spatio-temporal location. Particular tokens of the type will have a location, of course, but the type itself, a given piece of information$_{t}$, will not. Putting the point in the formal mode, `information$_{t}$' in both the quantum and classical settings is an abstract noun (in fact an abstract \textit{mass} noun), not a concrete one.
This result may or may not come as a surprise. What is undoubted is that there has been confusion over it, particularly when the nature of quantum teleportation has been up for discussion (see Section~\ref{understanding teleportation}).

The realisation that quantum information$_{t}$ is not a substance and is not part of the spatio-temporal contents of the world might lead on naturally to the conclusion that it therefore does not exist \textit{at all}; that there is no such thing as quantum information$_{t}$. This indeed was the conclusion of \citet{duwell:2003} although he has since retreated from this position to one closer to that advocated here \citep{duwell:qinfoexists}. The negative conclusion might be termed \textit{nihilism} about quantum information$_{t}$.   

Adopting a nihilist position, however, would appear to be an over-reaction to the fact that information$_{t}$ is not a material thing. As we have seen, quantum information$_{t}$ is what is produced by a quantum information$_{t}$ source. This will be an abstractum (type), but there is no need to conclude thereby that it does not exist. Many abstracta are very often usefully said to exist. To appreciate the point it is perhaps helpful to compare with a famous example of a non-existing substance.

So take `caloric'. This term was thought to refer to a material substance, one responsible for the thermal behaviour of various systems, amongst other things. But we found out that there was no such substance. So we say `Caloric does not exist'. But we also know now that there is no such substance as quantum information$_{t}$: why should we not therefore say `Quantum information does not exist'? 

The reason is that the two cases are entirely disanalogous, as the oddity of the phrasing in the previous sentence should immediately alert one to. The r\^{o}le of `caloric' was as a putative substance referring term; semantically it was a concrete noun, just one that failed to pick out any natural kind in this world. By contrast `information$_{t}$' was always an \textit{abstract} noun. It's r\^{o}le was \textit{never} that of referring to a substance. So it's not that we've discovered that there's no such substance as quantum information$_{t}$ (a badly formed phrase), but rather that attention has been drawn to the type of r\^{o}le that the term `information$_{t}$' plays. And this is not one of referring to a substance, whether putatively or actually. So unlike the case of caloric, where we needed to go out into the world and discover by experiment whether or not there is a substance called `caloric', we know from the beginning that the thought that there might be a substance called `information$_{t}$' is misbegotten, based on a misconception of the r\^{o}le of the term.

At this stage a further point must be addressed. One might be discomfited by my earlier comment that many abstracta are often usefully said to exist. Isn't this an area of some dispute? Indeed, wouldn't nominalists precisely be concerned to deny it? As it happens, though, the purposes of my argument may happily be served without taking a stand on such a contentious metaphysical issue. The point can be made that `information$_{t}$' is an abstract noun and that it therefore plays a fundamentally different r\^{o}le from a substance referring term; that it would be wrong to assert that quantum information$_{t}$ does not exist on the basis of recognising that quantum information$_{t}$ is not a substance; without having to take a stand on the status of abstracta. In fact all that is required for our discussion throughout is a very minimal condition concerning types that comes in both nominalist and non-nominalist friendly versions.

The non-nominalist version says the following: a piece of information$_{t}$, quantum or classical will be a particular sequence of states, an abstract type. What is involved in the type existing? Minimally, a sufficient condition for type existence will be that there be facts about whether particular concrete objects would or would not be tokens of that type. (Notice that this minimal condition needn't commit one to conceiving of types as Platonic objects). The nominalist version takes a similar form, but simply asserts that talk of type existence is to be paraphrased away as talk of the obtaining of facts about whether or not concrete objects would or wouldn't be instances of the type.

\subsubsection{A special case}

Having argued against the nihilist view and adressed possible nominalist concerns, we should close this section of the discussion by noting that there remains one special case in which it would seem to be correct to assert that quantum information$_{t}$ does not exist, the discussion so far notwithstanding.

Suppose one denied that there were facts about what quantum states systems possessed, or about what quantum operations devices implement. Then there will be no fact about what the output of a quantum source is, so there will be no fact about whether the systems produced are or are not an instance of any relevant type. In this event, it would be appropriate to maintain that quantum information$_{t}$ does not exist, as even the minimal criterion just given will not be satisfied. But does anyone hold this view of quantum mechanics? Yes: it is `quantum Bayesianism' as advocated by Caves, Fuchs and Schack (see, e.g., \citet{fuchs:only}) which we will be discussing in due course. For the quantum Bayesian, therefore, and perhaps only for them, it would be correct to say that quantum information$_{t}$ does not exist.

\subsection{Application: Understanding Teleportation}\label{understanding teleportation}

Why is it helpful to highlight the logico-grammatical status of information$_{t}$ as an abstract noun? In short, because the matter has given rise to confusion; and nowhere more so than in discussion of entanglement-assisted communication. One of the claims of \citet{grammar} is that failure to recognise that information$_{t}$ is an abstract noun is a necessary condition for finding anything conceptually problematic in teleportation, as so many have.

Here's how the story goes. The puzzles that teleportation presents cluster around two central questions. First, how is \textit{so much} information$_t$ transported in the protocol. And second, most pressingly, just \textit{how} does the information$_t$ get from Alice to Bob? We will concentrate on the second here (see \citet{grammar} for further discussion of the first).

A very common view is expressed by \citet{jozsa:1998,jozsa:2003} and \citet{penrose:1998}. In their view, the classical bits used in the protocol evidently can't be carrying the information$_{t}$: two classical bits are quite insufficient to specify the state teleported and in any case the bit values are entirely independent of the identity of the state. Therefore the entanglement shared between Alice and Bob must be providing the channel down which the information$_{t}$ travels. They conclude that in teleportation, an indefinitely large, or even infinite amount of information$_{t}$ travels backwards in time from Alice's measurement to the time at which the entangled pair was created, before propagating forward in time from that event to Bob's performance of his unitary operation and the attaining by his system of the correct state. Teleportation seems to reveal that entanglement has a remarkable capacity to provide a hitherto unsuspected type of information channel, one which allows information$_{t}$ to travel backwards in time; and a very great deal of it at that. It seems that we have made the discovery that quantum information$_{t}$ is a type of information$_{t}$ with the striking, and non-classical, property that it may flow backwards in time.   

The position is summarized succinctly by Penrose:

\begin{quote}
How is it that the \textit{continuous} ``information" of the spin direction of the state that she [Alice] wishes to transmit...can be transmitted to Bob when she actually sends him only two bits of discrete information? The only other link between Alice and Bob is the quantum link that the entangled pair provides. In spacetime terms this link extends back into the past from Alice to the event at which the entangled pair was produced, and then it extends forward into the future to the event where Bob performs his [operation].

Only \textit{discrete} classical information passes from Alice to Bob, so the complex number ratio which determines the specific state being ``teleported" must be transmitted by the \textit{quantum} link. This link has a channel which ``proceeds into the past" from Alice to the source of the EPR pair, in addition to the remaining channel which we regard as ``proceeding into the future" in the normal way from the EPR source to Bob.There is no other physical connection.
(\citet[p.1928]{penrose:1998})
\end{quote}

But this is a very outlandish picture. Is it really justified? \citet{dh} think not. They provide an analysis (based on a novel unitary, no-collapse picture of quantum mechanics) according to which the bits sent from Alice to Bob do, after all, carry the information$_{t}$ characterizing the teleported state. The information$_{t}$ flows from Alice to Bob, hidden away, unexpectedly in Alice's seemingly classical bits \footnote{The details of Deutsch and Hayden's approach and the question of what light it might shed on the notion of quantum information$_{t}$ is studied in detail in \citet{nifpaper} and \citet{dhshort}.}.

Trying to decide how the information$_{t}$ is transmitted in teleportation thus presents us with some hard questions. It looks like we have a competition between two different ontological pictures, one in which information$_{t}$ flows backwards, then forwards in time; the other in which the information$_{t}$ flows more normally, but hidden away inaccessibly in what we thought were classical bits. Perhaps we ought also to entertain the view that the information$_{t}$ just jumped non-locally somehow, instead. But what might that even mean?

The correct way out of these conundrums is to reject a starting assumption that they all share, by noting that there is something bogus about the question `How does the information$_{t}$ get from Alice to Bob?' in the first place.

Focus on the appearance of the phrase `the information$_{t}$' in this question. Our troubles arise when we take this phrase to be referring to a particular, to some sort of substance (stuff), perhaps, or to an entity, whose behaviour in teleportation it is our task to describe. This is the presumption behind the requirements of locality and continuity of information$_{t}$ flow that all of Jozsa, Penrose, Deutsch and Hayden apply in their various ways; and why it looks odd to think alternatively of the information$_{t}$ just jumping non-locally from Alice to Bob: things don't behave like that, we are inclined to think. All these approaches share the idea that information$_{t}$ is a kind of thing and that we need to tell a story about how this thing, denoted by `the information$_{t}$', moves about.

But when we recognise that `information$_{t}$' is an abstract noun, this pressure disappears. `The information$_{t}$' precisely does not refer to a substance or entity, or any kind of material thing at all; \textit{a fortiori} it is not something about which we can intelligibly ask whether \textit{it} takes a spatio-temporally continuous path or not. (By contrast, it  remains perfectly intelligible to ask the quite different question whether, in a given protocol, information$_{t}$ is transmitted \textit{by processes} that are spatio-temporally continuous.) Since `the information$_{t}$' does not introduce a particular, the question `How does the information$_{t}$ get from Alice to Bob?' cannot be a request for a description of how some thing travels. If it has a meaning, it is quite another one. It follows that the locus of our confusion is dissolved.

The legitimate meaning of `How does the information$_{t}$ get from Alice to Bob?', then, is just this: it is a roundabout way of asking what physical processes are involved in achieving the protocol. The end of the protocol is achieved when Bob's system is left in the same state as the one initially presented to Alice. That is what it is for the quantum information$_{t}$ to have been transmitted. We may then ask what physical processes were responsible for this; and the question will have a straightforward answer, \textit{although not one independent of your preferred interpretation of quantum mechanics}. You pay your money and you take your choice of the alternative, clear-cut, answers. See \citet[\S 5]{grammar} for a description in each of a variety of popular interpertations.

So while there can remain a source of disagreement about the physical processes involved in teleportation, co-extensive with disagreement over favoured interpretation of quantum mechanics, there is no longer any distinctive \textit{conceptual} puzzle left about the protocol. Once it is recognised that `information$_{t}$' is an abstract noun, it is clear that there is no further question to be answered regarding how information$_{t}$ is transmitted that goes beyond providing a description of the processes involved in achieving the end of the protocol. One doesn't face a double task consisting of a) describing the physical processes \textit{by which} information is transmitted, followed by b) tracing the path of a ghostly particular, information. There is only task (a).

The point should not be misunderstood: the claim is not that there is no such thing as the transmission of information$_{t}$, but simply that one should not understand the transmission of information$_{t}$ on the model of transporting potatoes, or butter, say, or piping water. 

Notice, finally, that the lesson developed here regarding teleportation applies equally in the case of superdense coding. There the source of puzzlement was how Alice could encode two classical bits into the single qubit she sends to Bob, given that the qubit she sends surely has to contain the information. But we should simply reject this latter premise, as it relies on the incorrect `thing' model of information$_{t}$.

\subsection{Summing up}\label{summing}

In this section we have seen how a straightforward explanation of what quantum information$_{t}$ is may be given; and seen moreover that there are very close links to the classical concept, despite Jozsa's misgivings we noted earlier. It is certainly true that quantum and classical information$_{t}$ differ in the types of sequence type that are involved---the quantum case requiring the richer structure of sequences of quantum states---but this does not preclude the two notions of information$_{t}$ from falling under a single general heading, from being, as advertised, species of a single genus. 

The crucial steps in the argument were, first, formulating the general definition of what information$_{t}$ is: that which is produced by a source that is required to be reproducible at the destination; and second, noting that the pertinent sense of `what is produced' is that which points us to the sequence types and not to the tokens. As a corollary we found that `information$_{t}$' is an abstract noun and therefore that neither classical nor quantum information$_{t}$ are parts of the material contents of the world.

Does this conclusion deprive quantum information$_{t}$ theory of its subject matter? Indeed not. It's subject matter in the abstract may be conceived of as the study of the structural properties of pieces of quantum information$_{t}$ (various sequences of quantum states and their possible transformations); and it's subject matter in the concrete may be conceived of as the study of the various new types of physical resources that the theory highlights (qubits and shared entanglement) and what may be done with them.

But finally, what bearing does all this have on the sorts of philosophical issues we noted in the introduction? We have seen the importance of being straight on the status of information$_{t}$ in understanding what is going on in teleportation. Two other things also follow quite directly, it seems. It is often claimed to be an important ontological insight deriving from, or perhaps driving, the success of quantum information$_{t}$ theory that `Information is Physical' \citep{landauer:1996}. Exactly what the role of this slogan might be deserves more detailed discussion \citep{thesis}, but things are quite clear on one reading, at least: it is simply a category mistake (we return to another reading later on). Pieces of information$_{t}$, quantum or classical, are abstract types. \textit{They} are not physical, it is rather their tokens that are. To suppose otherwise is to make the category mistake. Thus the slogan certainly does not present us with an ontological lesson. It might perhaps be thought that the purport of the lesson was actually supposed to be that we have made a discovery of a certain kind: that there really are physical instantiations of various pieces of quantum information (sequence types) possible in our world; and this need not have been so. Perhaps. But the force of \textit{this} lesson is surely limited: it should come as no surprise given that we already knew the world could be well described quantum mechanically.

The second point is this. As noted in the introduction, some have taken the development of quantum information$_{t}$ theory to support a certain kind of immaterialism (what might be called \textit{informational immaterialism}). Wheeler, for example, in his `It from Bit' proposal suggests that the basis of the physical world is really an immaterial one: `...that all things physical are information-theoretic in origin and this is a \textit{participatory universe}' \citep{wheeler}. This is an old metaphysical idea in the impressive modern dress of the most up-to-date of theories. But is such a view \textit{really} supported by the successes of quantum information$_{t}$ theory? It would seem not. 

We have seen that pieces of information$_{t}$ are abstracta. To be realised they will need to be instantiated by some particular token or other; and what will such tokens be? Unless one is \textit{already} committed to immaterialism for some reason, these tokens will be material physical things. So even if one's fundamental (quantum) theory makes great play of information$_{t}$, it will not thereby dispense with the material world. One needs the tokens along with the types. Thus we may safely conclude that immaterialism gains not one whit of support from the direction of quantum information$_{t}$ theory.

\section{The physical side of the theory of computation}\label{physicalside}

Quantum computation has presented a number of conceptual issues (see, e.g., \citet{deutsch:1985,FoR}, \citet{deutsch:etal:1999}, \citet{timpson:2004}). Here we shall highlight two. First, where does the computational speed-up come from in quantum computers? Second, what happens to the Church-Turing hypothesis in this context?

\subsection{Speed-up}

We have good reason to believe that quantum computers can be more efficient than classical ones: there is no known efficient classical algorithm for factoring, but there is a quantum one. It is interesting to ask where this speed-up comes from for at least two reasons. The first is a practical reason: If we had a better understanding of what was distinctively \textit{quantum} about quantum computation---the feature that allows the speed-up---then we would be better placed to develop further interesting quantum algorithms. The second, related, idea is more philosophical: understanding where the speed-up comes from would give us another handle on what the fundamental differences between classical and quantum systems are. Classical systems won't allow us to compute certain functions efficiently: what are the crucial differences that allow quantum systems to do so?

It is natural, although not wholly uncontroversial, to view the property of entanglement as the main source of the exponential speed-up given by quantum algorithms such as Shor's \citep{jozsa:1998,ekertjozsa:1998,jozsa:2000,jozsa?}. Ekert and Jozsa make the point that it cannot just be \textit{superposition} on its own that does the job, as classical systems that allow superpositions and thereby have vector spaces as their state space\footnote{Waves on strings would be an example---to get a finite dimensional state space, imagine confining yourself to the two lowest energy modes for each string. (Such a system is not a bit, of course, as there are a continuous number of distinct states given by superpositions of these two modes.)} would not allow speed-up. The crucial point seems to be how the state spaces for individual systems compose: classical vector space systems compose by the direct sum\footnote{The direct sum $V_{1}\oplus V_{2}$ of two vector spaces $V_{1}$, $V_{2}$, is a vector space composed of elements $f = \langle f_{i},f_{j}\rangle$, $f_{i}\in V_{1}$, $f_{j} \in V_{2}$; an ordered pair of elements of $V_{1}$ and $V_{2}$. If $\{g_{i,j}\}$ represents a basis for $V_{1,2}$ respectively, then a basis for $V_{1}\oplus V_{2}$ will be given by $\{\langle g_{i},\mathbf{0}\rangle, \langle \mathbf{0}, g_{j}\rangle \}$, hence $\mathrm{dim}V_{1}\oplus V_{2}= \mathrm{dim}V_{1} + \mathrm{dim}V_{2}$.} of the individual systems' state spaces (so $N$ 2-dimensional systems composed would have a dimensionality of 2$N$) whereas quantum state spaces compose by the tensor product (so the dimension of $N$ qubits is $2^{N}$) giving rise to entanglement.

However, even if we grant that entanglement plays a, or perhaps \textit{the}, crucial r\^{o}le, it is still possible to ask quite what the mechanism is. A popular answer has been in terms of parallel processing: we ought to think of the evolution of a quantum computer as a large number of distinct simultaneous computations. Indeed it has sometimes been suggested that the possibility of quantum computation provides resounding support for a Many Worlds view of quantum mechanics, as a way of understanding this parallel processing. Deutsch puts the point in characteristically forthright terms:
\begin{quote}
When a quantum factorization engine is factorizing a 250-digit number, the number of interfering universes will be of the order of $10^{500}$...
To those who still cling to a single universe world-view, I issue this challenge: \textit{explain how Shor's algorithm works}. I do not merely mean predict that it will work...I mean provide an explanation. When Shor's algorithm has factorized a number using $10^{500}$ or so times the computational resources that can be seen to be present, where was the number factorized? There are only about $10^{80}$ atoms in the entire visible universe, an utterly miniscule number compared with $10^{500}$. So if the visible universe were the extent of physical reality, physical reality would not even remotely contain the resources required to factorize such a large number. Who did factorize it, then? How, and here, was the computation performed? \citep[pp.216--7]{FoR}
\end{quote}        

But this rhetorical challenge is a plea on behalf of a fallacy; what can be called the \textit{simulation fallacy} \citep{grammar}: the fallacy of reading off features of a simulation as real features of the thing simulated, with no more ado. In this case, reading features of what would be required to provide a classical simulation of a computation as features of the computation itself. Deutsch assumes that a computation that would require a very large amount of resources if it were to be performed classically should be explained \textit{as} a process that consists of a very large number of computations, in Everettian parallel universes. But the fact that a very large amount of classical computation might be required to produce the same result as the quantum computation does not entail that the same amount of resources are required by the \textit{quantum} computer, or that the quantum computation consists of a large number of parallel classical computations. One can insist: why, after all, should the resources be counted in classical terms, to begin with? 
See \citet{steane} for further criticism of Deutsch's notion of parallel processing. (\citet{horsman} defends the intelligibility, if not the ineluctability, of the Many-Worlds analysis.)

The question of what classical resources would be required to simulate various quantum goings-on is a crucial idea in quantum information theory, but only for its pragmatic significance: it's a guide to possible new better-than-classical prtocols. It is by no means a guide to ontology.

Some recent theoretical developments shed further doubt on the parallel processing idea.

\paragraph{One-way computation} 

One-way quantum computation, also known as \textit{measurement-based} or \textit{cluster state} computation \citep{raussendorfbriegel:2001,raussendorfbrownebriegel:2003} is a very significant development for the practical implementation of quantum computation (see \citet{brownebriegel:review} for an introduction). In the standard quantum circuit model, a register of qubits is prepared in an initial, separable, computational basis state, which is then unitarily evolved by the action of the required sequence of gates on the qubits, typically into a complicated superposed entangled state, before perhaps ending with a measurement in the computational basis to read the result out. Different computations will take the register through different sequences of superposed entangled states with different unitary evolutions. By contrast, in one-way computing, a computation will begin with a network of qubits ready prepared in a particular kind of richly entangled state (a \textit{cluster} or \textit{graph} state); and different computations can start with the same state. The computation then proceeds by a sequence of measurements on single qubits and classical communication alone. There is no unitary evolution. Different algorithms will correspond to different sequences of one qubit measurements, where the basis in which a given measurement will be performed typically depends on the results of preceding measurements. It turns out that this system is easier to implement than the circuit model (no one or two qubit \textit{gates} are needed and no two qubit measurements: two qubit operations are the really tricky ones to achieve controllably) and it is considerably closer to current experimental capabilities. While standard quantum computation is reversible (up to any final measurement, at least), the one-way model is not (hence the name). The measurements at each step are irreversible and degrade the initial entanglement of the starting cluster state.

The point to take from this (as a number of people have emphasised, e.g. \citet{steane}) is that there is nothing in the one-way model of computation that looks like the parallel processing story; there are no linearly evolving parallel paths, as there is no unitary evolution. There is just a sequence of measurements banging on a large entangled state; the same state for different computations. Given that the one-way model and the circuit model are provably equivalent in terms of computational power, it follows that parallel processing cannot be the essence of quantum computational speed-up\footnote{A caveat. As far as I know no-one has yet attempted a description of one-way computing in fully unitary no-collapse quantum mechanics, i.e., where the measurements would be analysed quantum mechanically too. It's conceivable that such an analysis would reveal closer links to the circuit model than is currently apparent, although this is perhaps unlikely. Either way, the result would be of interest.}.

\paragraph{Bub's geometrical formulation} A more tentative, but nonetheless suggestive thought is this. Recently \citet{bub:algorithms} has provided a geometrical way of thinking about certain quantum algorithms that shows how apparently rather different looking algorithms, in particular, Deutsch's original XOR algorithm \citep{deutsch:1985} and Shor's algorithm, can be seen to exploit the same quantum mechanical fact in their operation: the fact that it is possible in quantum mechanics to compute the value of a disjunction without computing the values of the individual disjuncts. On this way of looking at things, rather than a quantum algorithm computing all the values at once---the parallelism idea---the point is that the algorithm is seen explicitly to avoid computing \textit{any} of the actual values of the function, these proving to be redundant for what the algorithm is aiming to achieve. What is particularly pertinent about Bub's analysis, though, is that it suggests that we may be asking the wrong question. The important point is that Shor's algorithm gives an exponential speed-up, whereas Deutsch's algorithm doesn't. So really what we thought we would have wanted was an analysis of these algorithms that makes them look \textit{different}, yet here they are illuminatingly cast as the same. So perhaps our question should not be `Why are quantum computers faster for some processes than classical ones?' but rather `Why is it that \textit{classical} computers are so \textit{slow} for some computations?'

\subsection{Whither the Church-Turing hypothesis?}

The study of quantum computation can, in some ways, be seen as a liberation for computer science. The classical Turing machine, abstractly characterised, had dominated theorising since its conception \citep{turing}. What the development of quantum computers showed was that \textit{just} focusing on abstract computational models, in isolation from the consideration of the physical laws govering the objects that might eventually have to implement them, can be to miss a lot. The progenitors of quantum computation realised that the question of what computational processes fundamental physics might allow was a very important one; and one which had typically been neglected in the purely mathematical development of computer science. One can argue that Turing's model of computing involved implicit classical assumptions about the kinds of physical computational processes there could be; hence his model was not the most general, hence Feynman's tongue-in-cheek remark \textit{a propos} Turing: `He thought he understood paper'\footnote{Cited by \citet[p.252]{FoR}.}. This is the line that \citet{deutsch:1985,FoR} explores. 

Thus quantum computers remind us that the theory of computing has two sides, the mathematical \textit{and} the physical; and that the interplay between them is important. We may miss things if our most general computational model does not in fact take into account all the possible kinds of physical process there are that might accomodate a computational reading; while a model that relies on processes that could not be physically implemented would not be an interesting one for practical purposes, perhaps would not even \textit{count} as a computational model. It turned out, of course, that quantum computers do not go wildly beyond Turing machines, they do not, for example compute the non-Turing computable; but they do instead raise important new questions in the rich theory of computational complexity\footnote{For an elementary discussion, see \citet{ultimate}, in more detail, \citet[Chpts. 3,4]{nielsen:chuang}.}. And the general point is well taken. For some, this is how the slogan `Information is Physical' is best read: as a needed corrective to computer science. Less ringing, perhaps, but more accurate, would be `Computers are Physical!'. 

In more strident application of the same point, it is significant to note that sensible proposals do exist for physically possible computations that would compute \textit{non-Turing} computable functions, e.g., \citet{hogarth}, \citet{shagrir:pitowsky} (although note the discussion in \citet{earman:norton}). 

Deutsch takes the lesson so far as saying that a new principle ought to replace the familiar Church-Turing hypothesis at the heart of the theory of computation, a physical principle which he calls the \textit{Turing Principle}:
\begin{quote}
Every finitely realizable physical system can be perfectly simulated by a universal model computing machine operating by finite means. \citep{deutsch:1985}
\end{quote}
Elsewhere I have argued that this is mistaken \citep{timpson:2004}. Here let us simply reflect on some crucial differences between two theses that are often confused. (\citet{copeland,copeland:encyclopedia} is exemplary in making such distinctions; see also \citet{pitowsky:2002}).

\paragraph{The Church-Turing Hypothesis} This is the claim, deriving from the seminal papers of \citet{church:1936} and \citet{turing} that \textit{the class of effectively calculable functions is the class of Turing machine computable functions}. This is a definition, or a stipulation, (in the material mode) of how the rough intuitive notion of effective calculability was to be formally understood. Given its definitional character, `hypothesis' is not really an apt name. It was important to provide such a definition of effective calculability in the 1930s because of the epistemological troubles in mathematics that drove Hilbert's formalist programme. The emphasis here is squarely on what can be computed by \textit{humans} (essential if the epistemological demands are to be met, see \citet[\S 3]{timpson:2004} and refs. therein) not anything to do with characterising the limits of machine computation.

\paragraph{The Physical Church-Turing thesis} This is a quite different thesis that comes in a variety of names and is often conflated with the Church-Turing hypothesis. It is the claim that \textit{the class of functions that can be computed by any physical system is co-extensive with the Turing computable functions}. Sometimes it comes in a stronger version that imposes some efficiency requirement: E.g., the efficiency of computation for any physical system is the same as that for a Turing machine (or perhaps, for a probabilistic Turing machine). This \textit{is} about the ultimate limits of machine computation. (Deutsch's Turing Principle is a thesis, directed towards the limits of physical computation, something along these lines; but where the concrete details of the Turing Machine have been abstracted away in the aim of generality.)\vspace{\baselineskip}

Notice that the kind of evidence that might be cited in support of these theses is quite different. In fact, since the first is a stipulation, it wouldn't make sense to offer evidence in support of its truth. All one can do is offer reasons for or against it as a good definition. The facts that are typically cited to explain its entrenchment are precisely of this form: one points to all the different attempts at capturing the notion of algorithm or of the effectively calculable: they all return the same class of functions (e.g. \citet[p.67]{cutland}). This tells us that Church and Turing did succeed in capturing the intuitive notion exceedingly well: we have no conflict with our pre-theoretic notions.

By contrast, the physical thesis is an empirical claim and consequently requires inductive support. It's truth depends on what you can get physical systems to do for you. The physical possibility of Malament-Hogarth spacetimes (and of the other elements required in Hogarth's protocol) for example, would prove it wrong. It's not clear how much direct or (more likely) indirect inductive support it actually possesses, certainly it should not be thought as deservedly entrenched as the Church-Turing hypothesis, although many are inclined to believe it. (Some admit: it's just a hunch.) What we do know is that quantum computation shows that the strong version, at least, is wrong (so long as no classical efficient factoring algorithm exists; and we believe none does).

Which of these two theses, if either, really lies at the heart of the theory of computation? In a sense, both: it depends what you want the theory of computation to be. If you are concerned with automated computing by machines and specifically with the ultimate limits of what you can get real machines to do for you, you will be interested in something like the physical version of the thesis, although one could clearly get along fine if it were false. If you are concerned with the notion of effective calculability and recursive functions, you will stick with the former thesis, the latter being largely irrelevant.

\subsubsection{Computational constraints on physical laws}

Some have been tempted to suggest that physical constraints on what can be computed should be seen as important principles governing physical theory. \citet{nielsen:ct} for example, argues that the physical Church-Turing hypothesis is incompatible with the standard assumption in quantum mechanics that a measurement can be performed for every observable one can construct (neglecting for present purposes dynamical constraints such as the Wigner-Araki-Yanase theorem \citep[pp.421--2]{peres}) and the thesis is also is incompatible with the possibility of unrestricted unitary operations. He conjectures that it is the physical Church-Turing thesis which should be retained and the required restrictions imported into quantum theory. Whether this is the correct conclusion to draw would depend on whether the inductive support for the physical thesis was greater than that accruing to quantum mechanics in its usual, unrestricted form. This seems questionable; although teasing out the evidence on either side would be an interesting task. A plausible default position might be that if one has in hand a well-confirmed and detailed physical theory that says that some process is possible, then that theory holds the trump card over a less specific generalisation covering the same domain. Consider the case of thermodynamics: this theory suggests that fluctuation phenomena should be impossible; kinetic theory suggests that they will happen---which one are you going to believe?\footnote{This leads us to an interesting general methodological issue: the default position just outlined looks plausible in some cases, but less so in others: consider the advent of Special Relativity in Einstein's hands. Perhaps in that case, though, one can point to specific defeating conditions that undermined the authority of the detailed theory in the domain in question.}   

Jozsa has presented another very interesting argument in similar vein (cf. \citet{jozsa?,jozsa:2003}). In his view, there is reason to think that computational complexity is a fundamental constraint on physical law. It is noteworthy that several different models of computation, very distinct physically---digital classical computing, analogue classical computing and quantum computing---share similar restrictions in their computing power: one can't solve certain problems in polynomial time. But this is for different reasons in the various cases. In the analogue case, for example, exponential effort would be needed to build sufficiently precise devices to perform the required computations, because it is very difficult to encode larger and larger numbers stably in the state of an analogue system. In the quantum case, one can see a restriction with measurement: if we could but read out all the results contained in a superposition then we would have enormous computational power; but we can't. 

Thus both analogue and quantum computation might appear to hold out the hope of great computing power, but both theories limit the ability to harness that power, while slight variations in the theories would allow one access to it\footnote{For an example of this in the quantum case, consider \citet{valentini:sqinfo} on sub-quantum information processing in non-equilibrium Bohm theory.}. This looks like a conspiracy on behalf of nature, or to put it another way, a case of homing in on a robust aspect of reality. Perhaps, then (the thought is) some general principle of the form `No physical theory should allow efficient solution of computational tasks of the class $x$' obtains. We might then use this as a guide to future theorising. However, it is unlikely that such a principle could sustain much commitment unless it were shown to mesh suitably with \textit{bona fide} physical principles. If one constructed a theory that was well-formed according to all physical desiderata one could think of, yet violated the computational complexity principle, it is implausible that one would reject it on those grounds alone.

\section{Foundations of QM}\label{foundations}

Whether advances in quantum information theory will finally help us to resolve our conceptual troubles with quantum mechanics is undoubtedly the most intriguing question that this new field holds out. Such diametrically opposed interpretational viewpoints as Copenhagen and Everett have drawn strength since its development. Copenhagen, because appeal to the notion of information has often loomed large in approaches of that ilk and a quantum theory of information would seem to make such appeals more serious and precise (more scientifically respectable, less hand-wavey); Everett, because the focus on the ability to manipulate and control individual systems in quantum information science encourages us to take the quantum picture of the world seriously; because of the intuitive appeal of Deutsch's many-worlds parallel processing view of algorithms; and most importantly, because of the theoretical utility of always allowing oneself the possibility of extending a process being studied to a unitary process on a larger Hilbert space. (This is known in the trade as belonging to the Church of the Larger Hilbert Space.) 

In addition to providing meat for interpretational heuristics, quantum information theory, with its study of quantum cryptography, error correction in quantum computers, the transmission of quantum information down noisy channels and so on, has given rise to a range of powerful analytical tools that may be used in describing the behaviour of quantum systems and therefore in testing our interpretational ideas. 

\subsection{Instrumentalism once more?}

As just mentioned, one strand in Copenhagen thought has always suggested that the correct way to understand the quantum state is in terms of information. One can see the (in)famous statement attributed to Bohr in just this light:
\begin{quote}
There is no quantum world. There is only an abstract physical description. It is wrong to think that the task of physics is to find out how nature \textit{is}. Physics concerns what we can say about nature. \citep{petersen:1963}
\end{quote}
Physics concerns what we can say about nature, not how things are; what we can say about nature---what information we have---is encoded in the quantum state. The state doesn't represent objective features of the world, it's just a means for describing our information. \citet{mermin:whose}, \citet{peierls:defence}, \citet{wheeler} and \citet{foundationalprinciple} have all been drawn to views of this nature. A canonical statement of the view is given by Hartle:
\begin{quote}
The state is not an objective property of an individual system but is that information obtained from knowledge of how a system was prepared, which can be used for making predictions about future measurements. \citep[p.709]{hartle:1968}
\end{quote}

With the flourishing of quantum information theory, which can indeed be seen, in a \textit{certain} sense, as taking quantum states to be information (cf. Section~\ref{whathow}) this view seems to acquire scientific legitimacy, even, perhaps, an attractive timeliness\footnote{The reader should draw their own conclusions about the validity of the train of thought involved, though. Notice that, when partaking in a quantum communication protocol, quantum states can be thought of as \textit{quantum} information; but wouldn't one want something more like \textit{classical} information when talking about Copenhagen-style measurement outcomes? Wouldn't one actually want information$_{e}$ rather than information$_{t}$ too? Reflect, also, on the discussion in Section~\ref{summing} for rebuttal of the idealist trend in the Bohr quotation.}.

There are some common objections to construing the quantum state as information from those of a more realist bent. Why, one might ask, if the quantum state is just information, should it evolve in accord with the Schr\"{o}dinger equation? Why should my state of mind, if you like, evolve in \textit{that} way? Yet we know the quantum state does (at least most of the time). Does it even make sense for cognitive states to be governed by dynamical laws? Or, one might be worried about where measurement outcomes are supposed to come from in this interpretation---measurement outcomes can't simply be information too, surely? Musn't they be part of the world? Neither of these are strong objections, though, both having simple answers. 

For the first, the reason that one's state of mind---the information one has that the quantum state represents---evolves in accord with the Schr\"{o}dinger equation (when it ought to), is that one subscribes to the laws of quantum mechanics. If a system is prepared in a certain way, then according to the theory, certain probabilities are to be expected for future measurement outcomes---this is what one comes to believe. If the system is then subject to some evolution, the theory tells you something specific: that what can be expected for future measurements will change, in a certain systematic way. It is because one is committed to quantum theory as descriptively accurate at the empirical level that one will update one's cognitive state appropriately. You know the rules for how states at $t_{1}$ are supposed to be related to states at $t_{2}$, so you assign them at those times accordingly.

As for the second, there is no requirement, on the view being adumbrated, that measurement outcomes be constituted by information (whatever that might mean) as there is no requirement that they be represented by a quantum state (e.g., we don't have to think of measurement pointer degrees of freedom taking on definite states as being constitutive of measurement outcomes). One can simply treat measurement outcomes as brute facts, happenings that will lead the experimenter to adopt certain quantum states in ways dictated by the theory, experimental context and their background beliefs.  

The real problem for the approach, indeed an insurmountable one, is presented rather by the following dilemma. 

The quantum state represents information? John Bell asked wisely: Information about what? \citep{bell:against} It seems that only two kinds of answer could be given:
\begin{enumerate}
\item Information about what the outcomes of experiments will be;
\item Information about how things are with a system prior to measurement, i.e., about hidden variables.
\end{enumerate}
Neither of these is satisfactory. The essential interpretive aim of construing the quantum state as information is to mollify worries about its odd behaviour (collapse, nonlocality). Such behaviour isn't troublesome if the state isn't describing physical goings-on. One argues: there's not really any \textit{physical} collapse, just a change in our knowledge; there's not really any \textit{nonlocality}, it's only Alice's knowledge of (information about) Bob's system that changes when she makes a measurement on her half of an EPR pair. But now suppose one opted for answer (2) to our question `Information about \textit{what}?', arguing that the information was about hidden variables. This would defeat the purpose of adopting this approach in the first place, as we all know that hidden variables are going to be very badly behaved indeed in quantum mechanics (nonlocality, contextuality). So our would-be informationist surely can't want this answer.

Turning then to the first answer, the trouble here is to avoid simply sliding into instrumentalism. An instrumentalist would assert that the quantum state is merely a device for calculating the statistics for measurement outcomes. How is the current view any different, apart from having co-opted the vogue term `information'? The point is, instrumentalism is not a particularly attractive or interesting interpretive option in quantum mechanics, amounting more to a refusal to ask questions than to take quantum mechanics seriously. It is scarcely the epistemologically enlightened position that older generations of physicists, suffering from positivistic hang-overs, would have us believe. If instrumentalism is all that appealing to information really amounts to, then there is little to be said for it. This shop-worn position is not made any more attractive simply by being re-packaged with modern frills.

A further fundamental problem for this approach is that `information' as it is required to feature in the approach, is a factive term. (I can't have the information that $p$ unless it is true that $p$.) This turns out to undermine the move away from the objectivity of state ascriptions it was the express aim of the approach to achieve. This matter is discussed in \citet[Chpt. 8]{thesis}. We may safely conclude that simply reading the quantum state in terms of information is not a successful move.

\subsection{Axiomatics}

If we are to find interesting work for the notion of information in approaching foundational questions in quantum mechanics we must avoid an unedifying descent into instrumentalism. A quite different approach is to investigate whether ideas from quantum information theory might help provide a perspicuous conceptual basis for quantum mechanics by leading us to an enlightening axiomatisation of the theory. We have seen that strikingly different possibilities for information transfer and computation are to be found in quantum mechanics when compared with the classical case: might these facts not help us characterise how and why quantum theory has to differ from classical physics? The most powerful expression of this viewpoint has been presented by Fuchs and co-workers (cf. \citet{fuchs:paulian}). We shall briefly survey three approaches in this vein. 

\subsubsection{Zeilinger's Foundational Principle}

\citet{foundationalprinciple} adopts an instrumentalist view of the quantum state along with a phenomenalist metaphysics: physical objects are assumed not exist in and of themselves but to be mere constructs relating sense impressions. Of more interest, and logically separable, is Zeilinger's concern in this paper to provide an information-theoretic foundational principle for quantum mechanics. The hope is to present an intuitively straightforward principle that plays a key r\^{o}le in deriving the structure of the theory. Zeilinger suggests he has found it in the principle that:
\begin{quote}
\textbf{Foundational Principle:} \textit{An elementary system represents the truth value of one proposition.}
\end{quote}
This is also expressed as the claim that elementary systems carry only one bit of information. 

Elementary systems are those minimal components that are arrived at as the end result of a process of analysis of larger composite systems into smaller component parts. In fact the Foundational Principle comes out as a tautology in this setting, as elementary systems are defined as those which can be described by a single (presumably, elementary) proposition only. (Shades of Wittgenstein's \textit{Tractatus} here.) The claim is that the Foundational Principle is the central principle for understanding quantum mechanics and that it explains both irreducible randomness and the existence of entanglement: key quantum features. It turns out, however, that the principle won't do the job \citep{supposed}. 

To see why, let us first cast the principle in more perspicuous form. As Zeilinger intends by `proposition' something that represents an experimenal question, the principle is the claim: \textit{The state of an elementary system specifies the answer to a single yes/no experimental question}. Then the explanation offered for randomness in quantum mechanics is that elementary quantum systems cannot, given the Foundational Principle, carry enough information to specify \textit{definite} answers to all experimental questions that could be asked. Therefore, questions lacking definite answers must receive a random outcome; and this randomness must be irreducible because if it \textit{could} be reduced to hidden properties, then the system would carry more than one bit of information. Entanglement is explained as arising when all of the $N$ bits of information associated with $N$ elementary systems are used up in specifying \textit{joint} rather than individual properties, or more generally, when more of the information is in joint properties than would be allowed classically \citep{essence}. What goes wrong with both of these purported explanations, however, is that no attention has been paid to the structure of the set of experimental questions on individual and joint systems. But without saying something about this, the Foundational Principle has no power at all.

Consider: irreducible randomness would only arise when there are more experimental questions that can be asked of an elementary system than its most detailed (pure) state description could provide an answer for. But what determines how many experimental questions there are and how they relate to one another? Certainly not the Foundational Principle. The Foundational Principle doesn't explain why, having given the finest grained state description we can manage, experimental questions still exist that haven't already been answered by our specification of that state. Put bluntly, why isn't one bit enough? (Compare a classical Ising model spin---here the one bit we are allowed per system is quite sufficient to answer all experimental questions that could be asked.) If we assume the structure of the set of questions is quantum mechanical, then of course such questions exist. But we cannot assume this structure: it is what we are trying to derive; and in the absence of any argument why space for randomness exists, we cannot be said to have explained its presence.

The story with entanglement is similar. We would only have an explanation of entanglement if it were explained why it is that there \textit{exist} experimental questions concerning joint systems to which the assignment of truth values is not equivalent to an assignment of truth values to questions concerning individual systems. It is only if this is the case that there can \textit{be} more information exhausted in specifying joint properties than individual ones, otherwise the joint properties would be reducible to individual ones. What we want to know is why this is the case; but the Foundational Principle cannot tell us.

As it stands, the Foundational Principle is wholly unsuccessful. Might we be able to salvage something from the approach, however? Perhaps if we were to add further axioms that entailed something about the structure of the set of experimental questions, progress could be made. A possible addition might be a postulate \citet{rovelli:relational} adopts: \textit{It is always possible to acquire new information about a system}. One wouldn't be terribly impressed by an explanation of irreducible randomness invoking the Foundational Principle and this postulate, however, as it would look rather too much like putting the answer in by hand. But there might be other virtues of the system to be explored. \citet{grinbaum:2005} discusses another axiom of similar pattern to Zeilinger's Foundational Principle, from a quantum logical perspective. \citet{spekkens} in a very suggestive paper, presents a toy theory whose states are states of less than maximal knowledge---the finest grained state description the theory allows leaves as many questions about the physical properties of a system unanswered as answered. What is remarkable is that these states display much of the rich behaviour that \textit{quantum} states display and which we have become accustomed to thinking is characteristic of quantum phenomena. The thought is that if such phenomena arise naturally for states of less than complete information, perhaps quantum states also ought to be thought of in that manner. Adopting this approach whole-heartedly, though, we would have to run once more the gauntlet outlined above of answering what the information was supposed to be about. 

\subsubsection{The CBH theorem}

A remarkable theorem due to Clifton, Bub and Halvorson (the CBH theorem) \citep{CBH} fares considerably better than Zeilinger's Foundational Principle. In this theorem, a characterisation of quantum mechanics is achieved in terms of three information-theoretic constraints (although it can be questioned whether all three are strictly necessary). The constraints are:
\begin{enumerate}
\item No superluminal information transmission between two systems by measurement on one of them;
\item no broadcasting of the information contained in an unknown state; and
\item no unconditionally secure bit-commitment.
\end{enumerate}
No broadcasting is a generalisation to mixed states of the no-cloning theorem \citep{nobroadcasting}. A state $\rho$ would be broadcast if one could produce from it a pair of systems $A$ and $B$ in a joint state $\tilde{\rho}_{AB}$ whose reduced states are both equal to $\rho$. This can obtain even when $\tilde{\rho}_{AB}\neq \rho\otimes\rho$, so long as $\rho$ is not pure. States can be broadcast \textit{iff} they commute. Arguably, no-broadcasting is a more intrinsically quantum phenomenon than no-cloning, because overlapping classical probability distributions cannot be cloned either, but they can be broadcast \citep{fuchs:thesis}.

Bit-commitment is a cryptographic protocol in which one party, Alice, provides another party, Bob, with an encoded bit value (0 or 1) in such a way that Bob may not determine the value of the bit unless Alice provides him with further information at a later stage (the `revelation' stage) yet in which the information that Alice gives Bob is nonetheless sufficient for him to be sure that the bit value he obtains following revelation is indeed the one Alice committed to originally. It turns out that this is a useful cryptographic primitive. A protocol is insecure if either party can cheat---Alice by being free to chose which value is revealed at revelation, or Bob by learning something about the value before revelation. Classically, there is no such protocol which is unconditionally secure. It was thought for a time that quantum mechanics might allow such a protocol, using different preparations of a given density matrix as a means of encoding the bit value in such a way that Bob couldn't determine it, but it was realised that Alice could always invoke a so-called EPR cheating strategy in order to prepare whichever type of density matrix she wished at the revelation stage \citep{lochau,mayers}. Instead of preparing a single system in a mixed state to give to Bob, she could present him with half of an entangled pair, leaving herself free to prepare whichever mixture she wished later. (See \citet{bub:commitment} for a detailed discussion.) We shan't dwell on bit-commitment, however as, arguably, it is a redundant condition in the CBH theorem (see \citet[\S 9.2.2]{thesis}). 

Finally we should note that the theorem is cast in the context of $C^{*}$-algebras, which CBH argue is a sufficiently general starting point as $C^{*}$-algebras can accomodate both quantum and classical theories\footnote{A $C^{*}$-algebra is a particular kind of complex algebra (a complex algebra being a complex vector space of elements, having an identity element and an associative and distributive product defined on it). A familiar example of a $C^{*}$-algebra is given by the set of bounded linear operators on a Hilbert space; and in fact any abstract $C^{*}$-algebra finds a representation in terms of such operators on some Hilbert space. One defines a state on the algebra, which is a positive, normalized, linear functional that can be thought of as ascribing expectation values to those elements of the algebra that represent observable quantities.}. The theorem states that any $C^{*}$-algebraic theory satisfying the information-theoretic constraints will be a quantum theory, that is, will have a non-commuting algebra of observables for individual systems, commuting algebras of observables for spacelike separated systems, and will allow entanglement between spacelike separated systems. The converse holds too (\citet{hans:generalization} filled-in a final detail) so the conditions are necessary and sufficient for a theory to be quantum mechanical. 

It is interesting and indeed remarkable that such a characterisation of quantum mechanics can be achieved and it undoubtedly enrichens our understanding of quantum mechanics and its links to other concepts, as one would hope for from a worthwhile novel axiomatisation of a theory. But with that said, questions have been raised both about the scope of the theorem and about what direct light it sheds on the nature and origin of quantum mechanics.

On the question of scope, a number of people have enquired whether the $C^{*}$-algebraic starting point is quite so neutral as CBH assumed. Both \citet{smolin} and \citet{spekkens} provided examples of theories satisfying the information-theoretic constraints, yet palpably failing to add up to quantum mechanics. What their constructions lacked were aspects of the $C^{*}$-algebraic starting point the theorem assumes. But for this very reason, their constructions raise the question: just how much work is that initial assumption doing? Concrete examples of the restrictiveness of the $C^{*}$-algebraic starting point may also be given \citep[\S 9.2.2]{thesis}. The $C^{*}$-algebraic notion of state implies that expectation values for observables must be additive. However, ever since Bell's critique of von Neumann's no-hidden variables assumption, it has been recognised that this is an extremely restrictive assumption \citep{bell:1966}. Insisting on beginning with $C^{*}$-algebras automatically rules out a large class of possible theories: hidden variables theories having quantum-mechanical structures of observables. This sort of criticism also relates to work by Valentini on the behaviour of general hidden variables theories which allow the possibility of non-equilibrium (i.e., non-Born rule) probability distributions \citep{valentini:sqinfo,valentini:sqinfodethv}. In such theories, empirical agreement with ordinary quantum mechanics is merely a contingent matter of the hidden variables having reached an equilibrium distribution. Out of equilibrium, markedly non-quantum behaviour follows, specifically, the possibility of instantaneous signalling and the possibility of distinguishing non-orthogonal states: two of the three information-theoretic conditions will be violated. From this perspective, the principles are not at all fundamental, but are accidental features of an equilibrium condition.

\paragraph{Interpretive issues} However it is over what conclusions can be drawn from the CBH theorem about the nature of quantum mechanics that the greatest doubts lie. In the original paper, some pregnant suggestions are made:
\begin{quote}
The fact that one can characterize quantum theory...in terms of just a few information-theoretic principles...lends credence to the idea that an information theoretic point of view is the right perspective to adopt in relation to quantum theory...We...suggest substituting for the conceptually problematic mechanical perspective on quantum theory an information-theoretic perspective...we are suggesting that quantum theory be viewed, not as first and foremost a mechanical theory of waves and particles...but as a theory about the possibilites and impossibilities of information transfer. \citep[p.4]{CBH}
\end{quote}
The difficulty is specifying what this amounts to. Given that the information-theoretic axioms have provided us with the familiar quantum mechanical structure once more, it is difficult to see that any of the debate over how this structure is to be interpreted, whether instrumentally or realistically, whether Copenhagen, collapse, Bohm, Everett, or what-not, is at all affected. Thus it is unclear how the information-theoretic perspective (however that is to be cashed out) could impinge on the standard ontological and epistemological questions; arguably it does not \citep[pp.214--222]{thesis}.

\citet{CBH} suggest that their theorem may be seen as presenting quantum mechanics as a \textit{principle} theory, as opposed to a \textit{constructive} theory, and this is where its interpretive novelty is to lie. The principle/constructive distinction is due to Einstein. Thermodynamics is the paradigm principle theory, to be contrasted with a constructive theory like the kinetic theory of gases. Principle theories begin from some general well-grounded phenomenological principles in order to derive constraints that any processes in a given domain have to satisfy. Constructive theories build from the bottom up, from what are considered to be suitably basic (and simple) elements and the laws governing their behaviour, to more complex phenomena. Einstein self-consciously adopted the principle theory approach as a route to Special Relativity.   

There are two problems here. The first: Even if one were to agree that quantum mechanics might usefully be viewed as a principle theory, where the principles are information-theoretic, then this would not take us very far. It would tell us that systems have to have certain $C^{*}$-algebraic states and algebras of observables associated with them, on pain of violation of the principles. But adopting this approach does not constrain at all how these states and observables are to be understood. Yet the usual interpretive issues in quantum mechanics lie at just this level: how are we to understand how the formalism is to map onto features of reality (if at all)? Remaining silent on this is simply to fail to engage with the central conceptual questions, rather than to present a less problematic alternative. 

The second point is that drawing an analogy with Einstein's (wonderfully successful!) principle theory approach to Special Relativity backfires \citep{browntimpson:2006}. Einstein was quite clear that constructive theories were to be preferred to principle theories and that constructive theories were more explanatory. He only reached for a principle theory methodology to obtain the Special Theory of Relativity as a move of desperation, given the confused state of physics at the turn of the 20th century; and was always unhappy with central elements of his original formulation of the theory thereafter (see \citet{harveybook} for more detail on this and on constructive alternatives to Einstein's original formulation). Einstein's 1905 methodology was a case of pragmatism winning out over explanatory depth. It is hard to see that an analogous manouevre would serve any purpose now, given that we already \textit{possess} quantum theory; and that this theory, in its quotidian form and application, is clearly \textit{the} constructive theory for physics.

\subsubsection{Quantum Bayesianism}

The final approach we shall consider is the most radical---and for that reason, the most interesting---one so far. This is the quantum Bayesianism of Caves, Fuchs and Schack \citep{fuchs:paulian,cfs:definetti,fuchs:only,fuchs:compatibility,fs:unknown04}. (Here we concentrate on the position as advocated by Fuchs.)

The quantum Bayesian approach is characterized by its non-realist view of the quantum state: the quantum state ascribed to an individual system is understood to represent a compact summary of an agent's degrees of belief about what the results of measurement interventions on a system will be. The probability ascriptions arising from a particular state are understood in a purely subjective, Bayesian manner. Then, just as with a subjective Bayesian view of probability there is no right or wrong about what the probability of an event is, with the quantum Bayesian view of the state, there is no right or wrong about what the quantum state assigned to a system is\footnote{The fact that scientists in the lab tend to agree about what states should be assigned to systems is then explained by providing a subjective `surrogate' for objectivity, along the lines that de Finetti provided for subjective probability: an explanation why different agents' degrees of beliefs may be expected to come into alignment given enough data, in suitable circumstances \citep{cfs:definetti}.}.  
The approach thus figures as the terminus of the tradition which has sought to tie the quantum state to cognitive states, but now, importantly, the cognitive state invoked is that of belief, not knowledge. The quantum state does not represent information, on this view (despite the occasional misleading claim to this effect), it represents an individual agent's subjective degrees of belief about what will happen in a measurement.

Importantly, however, this non-realist view of the quantum state is not the \textit{end point} of the proposal, but merely its \textit{starting point}. The aim is for more than a new formulation of instrumentalism and for this reason, it would be misguided to attack the approach as being an instrumentalist one. 
Rather, the hope expressed is that when the correct view is taken of certain elements of the quantum formalism (\textit{viz.} quantum states and operations) it will be possible to `see through' the quantum formalism to the real ontological lessons it is trying to teach us. Fuchs and Schack put it in the following way:
\begin{quote}
[O]ne...might say of quantum theory, that in those cases where it is not just Bayesian probability theory full stop, it is a theory of stimulation and response \citep{fuchs:what,fuchs:paulian}. The agent, through the process of quantum measurement stimulates the world external to himself. The world, in return, stimulates a response in the agent that is quantified by a change in his beliefs---i.e., by a change from a prior to a posterior quantum state. Somewhere in the structure of those belief changes lies quantum theory's most direct statement about what we believe of the world as it is without agents. \citep{fs:unknown04}
\end{quote}
Given the point of departure of a Bayesian view of the state, and using techniques from quantum information, the aim is to winnow the objective elements of quantum theory (reflecting physical facts about the world) from the subjective (to do with our reasoning). Ultimately, the hope is to show that the mathematical structure of quantum mechanics is largely forced on us, by demonstrating that it represents the only, or, perhaps, simply the most natural, framework in which intersubjective agreement and empirical success can be achieved given the basic fact (much emphasized in the Copenhagen tradition) that in the quantum domain, it seems that the ideal of a detached observer may not be obtained.

One of the main attractions of this approach, therefore, is that it aims to fill-in an important lacuna associated with many views in the Copenhagen tradition: It is all very well, perhaps, adopting some non-realist view of the quantum formalism, but, one may ask, why is it that our best theory of the very small takes such a form that it needs to be interpreted in this manner? Why are we forced to a theory that does not have a straightforward realist interpretation? Why is this the best we can do? The programme of Caves, Fuchs and Schack sets out its stall to make progress with these questions, hoping to arrive at some simple physical statements which capture what it is about that world that forces us to a theory with the structure of quantum mechanics. 

Note, however, that although the aim is to seek a transparent conceptual basis for quantum mechanics, there is no claim that the theory should be understood as a principle theory. In further contrast to the CBH approach, rather than seeking to provide an axiomatisation of the quantum formalism which might be interpreted in various ways, the idea instead is to take one particular interpretive stance and see whether this leads us to a perspicuous axiomatisation. 

This approach is self-consciously a research programme: If we adopt this view of the quantum formalism, where does it lead us? The proof of the pudding will be in the eating. The immediately pressing questions the approach raises are whether adopting the Bayesian approach would force us to give up too much of what one requires as objective in quantum mechanics, and what ontological picture goes along with the approach. How ought we to conceive a world in which the quantum Bayesian approach is the right one to take towards our best fundamental theory? These are matters for further investigation. 

\section{Outlook}

We have traversed a wide theoretical landscape and dwelt on quite a number of issues. Some conclusions have been clear and others left open. Some have been positive: we have seen how one ought to understand the notion of quantum information$_{t}$, for example, and how this helps us understand information$_{t}$ flow in entanglement-assisted communication. Others have been negative: we have seen the dangers of crude attempts to argue that the quantum state is information, or that quantum parallelism is a good argument for the Everett interpretation. Some important issues have been barely touched on, others not at all. Let's close by indicating a few of these last.  

Perhaps the most important philosophical issue that we have not discussed directly here is the general question of what kind of r\^{o}le the concept of information$_{t}$ has to play in physics. We have established some matters relevant to this: that information$_{t}$ is not a kind of stuff, so introduction of the concept of quantum information$_{t}$ is not a case of adding to the furniture of the world; but we have not attacked the issue directly. What we would like to be able to do is answer the question of what kind of physical concept information$_{t}$ is. Is it a fundamental one? (Might there more than one way in which concepts can be physically fundamental? Probably.) Or is it an \textit{adventitious} one: of the nature of an addition from without; an addition from the parochial perspective of an agent wishing to treat some system information-theoretically, for whatever reason? In addressing this issue it would be extremely helpful to have detailed comparisons with physical concepts that usually are taken as fundamental (relatively unproblematically), such as energy, charge and mass. (Notice that `energy', `charge', `mass', are all abstract nouns too; property names in fact. How does `information$_{t}$' differ from these? It is not always a property name, for one thing.)

A related theme is that of principle versus constructive theories, one we have touched on briefly. With its focus on task-oriented principles (you can't clone, you can't broadcast, you can't teleport without an ebit of entanglement), quantum information theory perhaps provides some scope for re-assessing preference for constructive theories over principle theories. If `information$_{t}$' features heavily in those latter theories, perhaps this would be an argument that information$_{t}$ is indeed a fundamental concept. Against that, one faces the perennial concern whether principle theories can ever really be truly fundamental.

Related again is the entire sphere of entanglement thermodynamics (see \citet{clifton:subtleties} for an invitation). The principle of no-increase of entanglement under local operations and classical communication appears to be in some ways akin to the second law of thermodynamics. Various analogies have been developed between entanglement and thermodynamic quantities \citep{pleniovedral:etd,rohrlich:analogues,horodeckis:balance}. It is a pressing concern to establish what these analogies have to teach us about the nature of entanglement and whether they are more than merely formal.

Another noteworthy ommission is any discussion of the thermodynamics of information$_{t}$ processing. This is an important issue that bears on the question of what one is to make of the notion of information$_{t}$ physically, particularly in discussion of Maxwell's demon, and in modern treatments of the exorcism of the demon that appeal to Landauer's Principle.  

Finally, one might wish to consider more explicitly the methodological lessons that quantum information$_{t}$ theory presents. One such lesson, perhaps, is that it provides an example of a theory of rich vigour and complexity in fundamental physics which does not proceed by introducing new kinds of material things into the world: it does not postulate new fundamental fields, particles, aether or ectoplasm. What it does do is ask new kinds of questions, illustrating the fact that fundamental physics need not always progress by the successful postulation of new things, or new kinds of things, but can also progress by introducing new general frameworks of enquiry in which new questions can be asked and in which devices are developed to answer them. Thus quantum information$_{t}$ theory might be another example to set alongside anaytical mechanics in Butterfield's call for more attention on the part of philosophers of science to the importance of such general problem setting and solving schemes in physics \citep{butterfield:lawsmodels}.   

\section{Further reading}

Systematic presentations of quantum information theory are given by  \citet{nielsen:chuang,physicsofqi,preskill,bennettshor:1998}. \citet{qcoding} is a very instructive read, as are Shannon's original papers \citep{shannon} (although there are one or two aspects of Shannon's presentation that have promulgated confusion down the years, c.f. \citet{jos} and discussion in \citet[\S 2]{supposed}).\vspace{\baselineskip}

\noindent \citet{bub:review} provides in many ways a fine complement to the discussion presented here. For more detail on many of the arguments and contentions I've presented, see \citet{thesis} and \citet{timpsonbook}.\vspace{\baselineskip}

\noindent \citet{fuchs:paulian} is a pithy and instructive collection of meditations on what significance quantum information theory might hold for the foundations of quantum mechanics, including the inside story on the mischieviously  polemical \citet{fuchs:peres}. \citet{fuchs:what} gives important background on the development of the quantum Bayesian position while \citet{cfs:certainty} provides the clearest statement to date of some important points.\vspace{\baselineskip}

\noindent A promising approach to coming to understand quantum mechanics better by getting a grasp on where it is located in the space of possible theories which allow information processing of various kinds is given by \citet{barrett:generalized}, building on the work of Popescu, Rohrlich and Hardy.\vspace{\baselineskip}

\noindent \citet{LexReff} is a most useful collection of essays on Maxwell's demon. See \citet{earmannorton:exorcism1,earmannorton:exorcism2,maroney:thesis,bennett:notes,maroney:absence,norton:eaters,ladyman:landp} for further discussion.\vspace{\baselineskip}

\noindent Volume 34, number 3 of \textit{Studies in History and Philosophy of Modern Physics} (2003) is a special issue on quantum information and computation, containing a number of the papers that have already referred to, along with others of interest. \citet{zurek:1990} is a proceedings volume of intriguing and relatively early discussions of the physics of information. \citet{teuscher:2004} is a stimulating volume of essays on Turing, his work on computers and some modern developments.

\section*{Acknowledgements}

I would like to thank Jeremy Butterfield, Harvey Brown, Jeff Bub, Ari Duwell, Chris Fuchs, Hans Halvorson, Pieter Kok, Joseph Melia, Michael Nielsen and Rob Spekkens for useful discussion. Thanks also to the editor for his invitation to contribute to this volume and for his patience. 

\newpage
{\footnotesize


}
\end{document}